\documentclass[12pt]{iopart}

\usepackage{graphicx,xcolor}
\expandafter\let\csname equation*\endcsname\relax
\expandafter\let\csname endequation*\endcsname\relax
\usepackage{amsmath, amsthm, amssymb}
\usepackage{float}
\usepackage{subfig}
\usepackage{xcolor}
\usepackage{fancyhdr}
\UseRawInputEncoding

\begin{document}

\title{Influence of plasma shaping on the parity of core-localized toroidal Alfv\'{e}n eigenmode in an advanced tokamak configuration }

\author{Shiwei Xue}

\address{State Key Laboratory of Advanced Electromagnetic Technology, \\International Joint Research Laboratory of Magnetic Confinement Fusion and Plasma Physics, School of Electrical and Electronic Engineering,
		\\	Huazhong University of Science and Technology, Wuhan, 430074,
		China}

\author{Ping Zhu*}

\address{State Key Laboratory of Advanced Electromagnetic Technology, \\International Joint Research Laboratory of Magnetic Confinement Fusion and Plasma Physics, School of Electrical and Electronic Engineering,
		\\	Huazhong University of Science and Technology, Wuhan, 430074,
		China;
~\\
Department of Nuclear Engineering and Engineering Physics, 
\\University of Wisconsin-Madison, Madison,
Wisconsin, 53706, United States of America}
\ead{zhup@hust.edu.cn}

\author{Haolong Li}

\address{College of Sciences, Tianjin University of Science and Technology, Tianjin 300457, China}

\vspace{10pt}
\begin{indented}
\item[]\today
\end{indented}
\clearpage
\begin{abstract}

Toroidal Alfv\'{e}n eigenmodes (TAEs) and energetic particle modes (EPMs) can both be excited by energetic particles (EPs) from auxiliary heating and fusion-born $\alpha$ particles in a tokamak. Using the hybrid kinetic-MHD (HK-MHD) model implemented in the NIMROD code, we have demonstrated the excitation of these modes and their behaviors in an advanced tokamak configuration with {reversed magnetic shear in the core region.} {The} TAE/EPM predominantly exhibits odd parity {and anti-ballooning structure when the plasma assumes elongated non-circular 2D shaping. However, as the 2D plasma shaping becomes more circular with reduced elongation, the TAE/EPM mode parity eventually transitions to even along with the ballooning structure. Such a finding may explain the dominant mode parity of TAE/EPMs observed in an advanced tokamak configuration with any specific 2D plasma shaping.}
\end{abstract}

\vspace{2pc}

\submitto{\PPCF}

\pagestyle{fancy} 
\fancyhf{} 
\fancyhead[C]{\itshape {Plasma shaping and  parity of core-localized toroidal Alfv\'{e}n eigenmode}} 
\renewcommand{\headrulewidth}{0.0pt}
\fancyfoot[C]{\thepage} 
\section{Introduction}

In recent decades, both theoretical \cite{CHENG198521,chengLowShearAlfven} and experimental \cite{wongExcitationToroidalAlfven1991,wongReviewAlfvenEigenmode1999,zhuNonlinearModeCouplings2022,kimSuppressionToroidalAlfven2022,vallarExcitationToroidalAlfven2023,fitzgeraldStabilityAnalysisAlpha2023} studies {have demonstrated the excitation of} the TAE instability {in tokamaks} by energetic particles. Two distinct types of TAEs have been identified: global TAE and core-localized TAE. {The} global TAE occurs when a few poloidal harmonics exhibit comparable peak values, resulting in a mode structure that spans a substantial fraction of the plasma volume \cite{CANDY1996299}. In contrast, {the} core-localized TAE has a highly localized mode structure within a single TAE gap near {the rational surface where the safety factor} $q = m/n$,{with $m$ and $n$ being the poloidal and toroidal mode numbers of the TAE} which {can be} easily destabilized when the energetic ion density {profile} peaks {at the center of plasma} \cite{fuExistenceCoreLocalized1995}. In this case, there are typically two dominant poloidal harmonics, namely, $m$ and $m + 1$. And based on the signs of the two harmonics, core-localized TAEs are categorized into even TAEs and odd TAEs. The even mode, situated at the bottom end of the TAE gap, is formed by the coupling of poloidal harmonics with the same sign. In contrast, the odd mode, located at the top end of the TAE gap, exhibits opposite signs between its two poloidal components. {Previous theory predicts that the} 
existence of core-localized TAEs depends on the normalized background plasma pressure gradient, defined as $\alpha=-2\left(R q^2 / B^2\right) p^{\prime}$, being below a critical threshold. For the even mode, this critical value is approximately $\alpha_c^E \approx 3 \epsilon+2 s^2$, whereas for the odd mode, it is $\alpha_c^O \approx 3 \epsilon-2 s^2$, where $\epsilon$ is the inverse aspect ratio and $s$ is the magnetic shear{~\cite{10.1063/1.871537}}. Consequently, {in theory} the odd mode can only exist when the shear is sufficiently low, and its critical $\alpha$ is lower than that of the even mode~\cite{fuExistenceCoreLocalized1995,berkMoreCoreLocalized1995}. In addition to these equilibrium constraints, the excitation of odd TAEs is generally more challenging because of the finite orbit width effect~\cite{10.1063/1.871537,fuStabilityAnalysisToroidicityInduced1995}. Nevertheless, odd TAEs were first observed experimentally on JET~\cite{kramerObservationOddToroidal2004}. In subsequent numerical studies, they were identified in {EAST} simulations with the NIMROD code~\cite{nimrodteamNumericalStudyTransition2019}, and more recently, FAR3d simulations have reproduced odd TAEs in JET D-T discharges dominated by passing energetic particles~\cite{varelaAnalysisLinearNonlinear2025}.

In the past few years, the {physical and engineering} design of the CFETR \cite{wanPhysicsDesignCFETR2014,wanOverviewPresentProgress2017} has made substantial progress, which is proposed to bridge the {r}esearch and {d}evelopment gaps between  International Tokamak Experimental Reactor (ITER) \cite{gormezanoChapterSteadyState2007,kikuchiSteadystateTokamakResearch2012} and fusion DEMOnstration reactor (DEMO) \cite{zohmPhysicsGuidelinesTokamak2013}. {In the CFETR baseline scenarios,} {t}here are {large amount} of energetic ions ($100 keV - 1 MeV$) {generated during} NBI and RF heating, as well as {the} $3.5 MeV$ alpha particles produced in D-T reaction. {Thus} AEs can be easily excited by these EPs when their drive exceeds other damping mechanisms, such as {the} continuum damping, Landau damping, as well as {the} radiative damping. {To study the basic features of AEs in the CFETR steady-state scenario, various codes such as NOVA/NOVA-K \cite{yangLinearStabilityToroidal2017}, FAR3d \cite{varelaTheoreticalStudyAlfven2022}, and GEM \cite{renStabilityAlfvenEigenmodes2020,renInvestigationAlphaparticleTransport2024} have been used to analyze the EP-driven modes.}
 
In this work, the dominant existence of odd-parity TAEs/EPMS with an anti-ballooning structure in the CFETR-like baseline scenario is demonstrated using both the GTAW \cite{huNumericalStudyAlfven2014a} and NIMROD \cite{sovinecNonlinearMagnetohydrodynamicsSimulation2004, kimHybridKineticMHDSimulations2004, kimImpactVelocitySpace2008} codes. This is consistent with the previous theory prediction that the odd TAEs are more likely to appear in the advanced configurations with a zero magnetic shear region. Parametric scans of the minimum safety factor \(q_{\text{min}}\) and the EP $\beta$ fraction \(\beta_h\) indicate that these parameters do not significantly alter the mode structure or parity. However, when new equilibrium generated by the {EFIT}\cite{laoReconstructionCurrentProfile1985} and CHEASE \cite{lutjensAxisymmetricMHDEquilibrium1992,lutjensCHEASECodeToroidal1996a} are employed, it is found that reducing the plasma elongation gradually to a certain threshold induces a clear transition of the TAE from odd to even parity. This highlights the critical role of plasma shaping in determining the TAE/EPM parity in the advanced tokamak scenario that has not been captured by any previous theory.

The rest of this paper is organized as follows. Section \ref{simulation model} reviews the hybrid kinetic-MHD model implemented in the NIMROD code. The main CFETR parameters and profiles used in our simulations are detailed in Section \ref{simulation setup}. We present our simulation results in Section \ref{results}, demonstrating that while $q_{\text{min}}$ and $\beta_h$ have a negligible influence on the intrinsic mode properties, plasma shaping induces a clear transition between odd and even TAEs. Finally, a summary and discussion are provided in Section \ref{summary}.

\section{{Hybrid kinetic-MHD} model}

\label{simulation model}
For the HK-MHD model implemented in the NIMROD code, the background plasma and energetic ions follow using MHD equations and drift kinetic equations respectively \cite{sovinecNonlinearMagnetohydrodynamicsSimulation2004,kimHybridKineticMHDSimulations2004}. {In particular, the single-fluid ideal MHD equations are}
\begin{center}
	\begin{gather}
		\frac{\partial \rho}{\partial t}+\nabla \cdot(\rho \boldsymbol{V})=0 \\
		\rho\left(\frac{\partial \boldsymbol{V}}{\partial t}+\boldsymbol{V} \cdot \nabla \boldsymbol{V}\right)=\boldsymbol{J} \times \boldsymbol{B}-\nabla p_b-\nabla \cdot \boldsymbol{P}_h \\
		\frac{1}{\Gamma-1}\left(\frac{\partial p}{\partial t}+\boldsymbol{V} \cdot \nabla p\right)=-p \nabla \cdot \boldsymbol{V} \\
		\frac{\partial \boldsymbol{B}}{\partial t}=-\nabla \times \boldsymbol{E} \\
		\boldsymbol{J}=\frac{1}{\mu_0} \nabla \times \boldsymbol{B} \\
		\boldsymbol{E}+\boldsymbol{V} \times \boldsymbol{B}=0
	\end{gather}
\end{center}
where subscripts $b, h$ denote {the} bulk plasma and {the hot or} fast particles, $\rho, \boldsymbol{V}$ are the mass density and {the} velocity {of} bulk plasma, neglecting the contribution of fast particles, $p$ is the pressure of entire plasma, $p_b$ is the pressure of bulk plasma, {$\boldsymbol{P}_h$} is the pressure tensor of fast particles, and {$\Gamma$} is {the} ratio of specific heats. {The rest of the symbol definitions are conventional.}

In the above HK-MHD model, it is assumed that the density of fast species is much lower than that of bulk plasmas but the fast species pressure is on the order of the bulk plasma pressure, i.e. ${n_h} \ll n_b$ and ${\beta_h} \sim \beta_b$, and $\beta \equiv 2 \mu_0 p / B^2$ is the ratio of thermal energy to magnetic energy \cite{chengKineticmagnetohydrodynamicModelLowfrequency1991a}. In this approximation, we neglect the contribution of energetic particles to the center of mass velocity. If we take the center of the mass velocity of energetic ions to be zero, ${\boldsymbol{P}_h}$ in the momentum equation can be calculated from the velocity distribution function of energetic ions. The $\delta f$ PIC method is utilized to solve the drift kinetic equation for energetic particles {\cite{kimHybridKineticMHDSimulations2004}}.
\begin{equation}
	{ 
	\label{eq:drift1}
		 \dot{\boldsymbol{x}}=v_{\|} \hat{\boldsymbol{b}}+\frac{m}{e B^4}\left(v_{\|}^2+\frac{v_{\perp}^2}{2}\right)\left(\boldsymbol{B} \times \nabla \frac{B^2}{2}\right)+\frac{\boldsymbol{E} \times \boldsymbol{B}}{B^2}+\frac{\mu_0 m v_{\|}^2}{e B^2} \boldsymbol{J}_{\perp}} 
\end{equation}
\begin{equation}
	{ 
	\label{eq:drift2}
	m \dot{v}_{\|}=-\hat{\boldsymbol{b}} \cdot(\mu \nabla B-e \boldsymbol{E})}
\end{equation}
\noindent
{where $v_\perp$ ($v_\parallel$) is the velocity perpendicular (parallel) to the magnetic field, $\mu$ is the magnetic moment, $\hat{\boldsymbol{b}}=\boldsymbol{B}/B$ is the unit vector along the magnetic field, $m$ is the mass of the energetic particle, and $e$ is the electric charge.} {The individual terms in Eq.~\eqref{eq:drift1} correspond to the standard drift
	velocities in the drift–kinetic description. The first term,
	$v_{\|}\hat{\boldsymbol{b}}$, gives the parallel motion along the magnetic field.
	The second term,
	represents the combined curvature and $\nabla B$ drift. The third term corresponds to the 
	$\boldsymbol{E}\times\boldsymbol{B}$ drift. And the last term is the finite-pressure correction to the curvature and $\nabla B$ drifts, where $\boldsymbol{J}_{\perp} = 
	\boldsymbol{J}-\boldsymbol{J}\!\cdot\!\hat{\boldsymbol{b}}\,\hat{\boldsymbol{b}}$ \cite{kimImpactVelocitySpace2008}.} {Assume the phase space distribution function $f_h=f_{h0}+\delta f_{ h}$, where {$f_{h0}$} and {$\delta f_{ h}$} are the equilibrium and the perturbed distribution functions of
energetic particles, this gives} {$\boldsymbol{P}_h=\boldsymbol{P}_{h0}+\delta \boldsymbol{P}_h$}, where {$\boldsymbol{P}_{{h0}}$} is assumed isotropic, and {$\delta \boldsymbol{P}_{{h}}$} is defined as 
\begin{equation}
	\label{eq:delta p}
	\delta {\boldsymbol{P}_{{h}}}=\left(\begin{array}{ccc}
		\delta p_{\perp} & 0 & 0 \\
		0 & \delta p_{\perp} & 0 \\
		0 & 0 & \delta p_{\|}
	\end{array}\right)
\end{equation}
{where $\delta p_{\perp} = \int \mu B \delta f_h d^3 v$ {($\delta p_\parallel = \int v_{\|}^2 \delta f_h d^3 v$)} is the {stress tensor component} due to hot particle motions perpendicular {(parallel)} to the magnetic field \cite{kimImpactVelocitySpace2008}}. 

\section{{Numerical} setup}
\label{simulation setup}

{The simulation is based on a designed equilibrium for the CFETR steady-state scenario and the simulation domain enclosed with the last closed flux surface (LCFS) is represented using a 2D bicubic finite element mesh aligned with the equilibrium magnetic flux surfaces (Figure \ref{figure1}).}

\begin{center}
	[Figure 1 about here.]\\
\end{center}

{{The equilibrium has a} $q$ profile with shear reversal at a large {minor radial location} is adopted to {help achieve} the {desired} plasma performance.} Since the bootstrap fraction is not very high in the baseline scenario with a moderate $\beta_N$, the reversed shear setup requires a large off-axis current to be driven by external sources such as NBI, which can also introduce energetic particles.

The slowing down distribution is employed for such EPs \cite{kimImpactVelocitySpace2008}:
\begin{equation}
	f_0 = \frac{P_0 \exp \left( \frac{P_\zeta}{\psi_n} \right)}{\varepsilon^{3/2} + \varepsilon_c^{3/2}}
\end{equation}
where $P_0$ is the normalization constant, $P_\zeta=g \rho_{\|}-\psi_{\mathrm{p}}$ is the canonical toroidal momentum, $g=R B_\phi, \rho_{\|}=m v_{\|} / q B, \psi_{\mathrm{p}}$ is the poloidal flux, $\psi_{\mathrm{n}}=c \psi_0, \psi_0$ is the {total poloidal magnetic
	flux} and the parameter $c$ is used to match the spatial profile of the equilibrium, $\varepsilon$ is the particle energy, and $\varepsilon_{\mathrm{c}}$ is the critical slowing down energy 
\begin{equation}
	\varepsilon_{\mathrm{c}}=\left(\frac{3}{4}\right)^{2 / 3}\left(\frac{\pi m_{\mathrm{i}}}{m_{\mathrm{e}}}\right)^{1 / 3} T_{\mathrm{e}}
\end{equation}
with $m_{\mathrm{i}}$ being the ion mass, $m_{\mathrm{e}}$ the electron mass, and $T_{\mathrm{e}}$ the electron temperature. When $\varepsilon>\varepsilon_{\mathrm{c}}$, the slowing down of beam ions is mainly due to the collisions with background electrons, {whereas} the collisions with background ions {become} dominant when $\varepsilon<\varepsilon_{\mathrm{c}}$. {All other key parameters are also set up based on the designed CFETR scenario and can be found in Table \ref{table:parameters} \cite{zhouOptimizationsCFETRSteady2022}} .
\begin{table}[h!]
	\centering
	\caption{CFETR Main parameters}
	\label{table:parameters}
	\begin{tabular}{|l|c|}
		\hline
		\textbf{Parameter} & \textbf{Value} \\
		\hline
	    EPs' $\beta$ fraction,	$\beta_h$ & 0.43 \\
		Major radius, $R$ & 7.2 m \\
		Minor radius, $a$ & 2.2 m \\
		{T}oroidal magnetic field {at magnetic axis}, $B_0$ & 6.5 T \\
		\hline
	\end{tabular}
\end{table}

\section{{Calculation results and analyses of AE transitions}}
\label{results}
{We first calculate the} stability and structure of $n=1-6$ modes. We then focus on the $n=3$ mode, {identified} as an odd-parity EPM. An investigation into the effects of the $q_{\text{min}}$ and $\beta_h$ reveals that neither parameter significantly influences the mode's parity. Subsequently, we demonstrate {its} transition to an odd-parity TAE by reducing the background plasma $\beta$ to approximately one fifth of its initial value. Using CHEASE, we then show that a gradual reduction in plasma elongation on this new equilibrium leads to a clear transition {of the TAE} from odd-parity to even-parity.

\subsection{Initial calculation and mode identification}

\begin{center}
	[Figure 2 about here.]\\
\end{center}

\begin{center}
	[Figure 3 about here.]\\
\end{center}

\noindent For $\beta_h = 0.43$, both the frequency and growth rate exhibit a bell-shaped dependence on the toroidal mode number $n$ (Figure \ref{fig:fg}). The growth rate reaches its maximum at $n=3$, whereas the frequency attains its peak at $n=4$. {T}he frequencies of {$n=2$--$6$ modes are located in the continuum} close to $0.5 R_0/V_A$. {An exception is the $n=1$ branch, whose continuum, parity and spectral characteristics differ qualitatively from the $n=2$--$6$ modes; more details are provided in~\ref{app:a}.}  Attention is then directed to the mode identification of the $n=3$ case. Based on the analysis of the poloidal Fourier spectrum (PFS) result (Figure \ref{fig:pfs_n3_org}) together with the Alfv\'en continuum (Figure \ref{fig:continuum_n3}), the $n=3$ mode is identified as an EPM. Moreover, it is worth noting that Figure~\ref{fig:contour} shows the mode structures are predominantly anti-ballooning (odd parity) for almost all cases, which is consistent with previous theory that odd TAE/EPMs are more likely to exist in the presence of a $q_{\min}$ region with zero magnetic shear\cite{kramerObservationOddToroidal2004}. 
It should also be mentioned that, throughout this work, the PFS analysis retains only the two dominant poloidal harmonics with the largest amplitudes, so the plotted spectra represent the strongest coupled $m$ components of the mode.

\begin{center}
	[Figure 4 about here.]\\
\end{center}

\subsection{$q_{\text{min}}$ effects}
To assess the impact of the $q$ profile with reversed magnetic shear on Alfv\'en eigenmodes, the frequency, linear growth rate, and mode structure of the $n=3$ mode are evaluated for various values of the minimum safety factor $q_{\text{min}}$ . 

\begin{center}
	[Figure 5 about here.]\\
\end{center}

Figure~\ref{fig:freq_q} and \ref{fig:growth_q} show that both the mode frequency and growth rate undergo oscillatory variations as $q_{\min}$ is varied. In contrast, the contour plots of $V_{\psi}$ in Figure~\ref{fig:qmin215} and \ref{fig:qmin250} demonstrate that the spatial structure remains essentially unchanged, consistently retaining {the} anti-ballooning (odd parity) {pattern} across the entire range of $q_{\min}$. These results indicate that although $q_{\min}$ influences the quantitative values of frequency and growth rate, its effect on the mode parity is negligible.

\subsection{{\(\beta_h\) effects}}

To study the EP beta fraction effect of energetic particles, we also choose $n = 3$ mode as an example and vary {$\beta_h$}. The influence of the energetic particle fraction $\beta_h$ on the $n=3$ mode is summarized in Figure~\ref{fig:combined_beta}. 
\begin{center}
	[Figure 6 about here.]\\
\end{center}

As shown in Figures~\ref{fig:freq_beta} and \ref{fig:growth_beta}, an increase in $\beta_h$ significantly enhances the linear growth rate, while the mode frequency exhibits a slight reduction. In contrast, the contour plots of $V_{\psi}$ in Figures~\ref{fig:beta016} and \ref{fig:beta060} indicate that the spatial structure remains essentially unchanged, consistently retaining {the} anti-ballooning (odd parity) {pattern} across the examined range of $\beta_h$. These results demonstrate that although $\beta_h$ has a strong impact on the growth rate and a modest influence on the frequency, its effect on the mode parity is {also} negligible.

\subsection{Plasma shaping effects}
\begin{center}
	[Figure 7 about here.]\\
\end{center}

\noindent Up to this point, the even TAE, which in theory is supposedly at least equally possible to excite in the region with zero magnetic shear \cite{fuStabilityAnalysisToroidicityInduced1995}, has {appeared elusive} in the results presented above. To explore the conditions under which the even TAE may emerge, the background plasma {$\beta$} is first reduced to {about} one {fifth} of its original equilibrium value (Figure~\ref{pres2}), while the safety factor profile {in the core region} is kept as close as possible to that of the initial equilibrium (Figure~\ref{safe2}). New equilibrium is obtained using the {EFIT} code. For the EP fraction $\beta_h=0.43$, the Alfv\'en continuum (Figure~\ref{continuum_new_n3}), the 2D mode structure in poloidal plane (Figure~\ref{con_new_n3}), and the poloidal Fourier spectrum (Figure~\ref{pfs_new_n3}) reveal that the n=3 mode remains independent of the reduction of background plasma equilibrium pressure.

\begin{center}
	[Figure 8 about here.]\\
\end{center}

\begin{center}
	[Figure 9 about here.]\\
\end{center}

Next, the CHEASE code is employed to adjust the plasma shaping by varying the equilibrium elongation, denoted as $\kappa=b/a$, where $a$ and $b$ are the half horizontal and vertical diameters of the plasma poloidal cross-section, respectively. The elongation is varied from $\kappa=2.0$, which approximates the CFETR equilibrium, down to the circular limit of $\kappa=1.0$ (Figure~\ref{ellipse}). The contours and PFS results (Figure~\ref{fig:combined_ba}) clearly show that as $\kappa$ decreases from $2.0$ to $1.22$, the TAE undergoes a transition from odd-parity to even-parity. In other words, the mode structure changes from {the} anti-ballooning to a ballooning {pattern}. {It should be noted that this shaping-induced parity transition is only found in equilibria with weak or reversed magnetic shear. For configuration with strong positive shear only, this effect is absent, as noted in detail in the \ref{app:b}.}
\begin{center}
	[Figure 10 about here.]\\
\end{center}

\section{Conclusions and discussion}
\label{summary}
In this work, EPMS/AEs in the CFETR baseline scenario are investigated with the hybrid kinetic–-MHD module of NIMROD and the eigenvalue code GTAW, {along with the equilibrium codes} EFIT and CHEASE. In a weak reverse-shear equilibrium, both TAEs and EPMs can be destabilized across several toroidal mode numbers. In this configuration, the unstable modes preferentially exhibit odd parity in the poloidal plane, i.e., an anti-ballooning structure. Within the parameter ranges explored, moderate variations in the minimum safety factor $q_{\text{min}}$ and the energetic particle fraction $\beta_h$ do not qualitatively alter the mode parity. When the bulk plasma pressure is reduced, the $n=3$ branch transitions from an EPM to an odd-parity TAE. A further reduction of plasma elongation to $\sim\!1.22$ then drives a clear transition {of the TAE} from odd to even {parity}, i.e., from anti-ballooning to ballooning structure. {Further analysis in equilibrium with positive magnetic shear only indicates that such a transition is likely to be limited to the configurations with zero magnetic shear.}

Taken together, these results suggest that in advanced tokamak configurations relevant to burning plasmas, EP-driven modes (TAEs/EPMs) are prone to anti-ballooning (odd parity) structures under the combined effects of high pressure, weak reverse shear, and finite elongation. Conversely, decreasing pressure and reducing elongation toward circular cross sections tend to favor the even-parity TAEs even in the presence of zero magnetic shear. These findings provide important insights for the future design of advanced tokamaks.

\section*{Acknowledgements}

We are grateful for the supports from the NIMROD team. This work is supported by the National MCF Energy R\&D Program of China Grant No.~2019YFE03050004, {the Hubei International Science and Technology Cooperation Project Grant No.~2022EHB003}, and the U.S. Department of Energy Grant No.~DE-FG02-86ER53218. The computing work in this paper is supported by the Public Service Platform of High Performance Computing by Network and Computing Center of HUST{, and this research used resources of the National Energy Research Scientific Computing Center, a DOE Office of Science User Facility supported by the Office of Science of the U.S. Department of Energy under Contract No. DE-AC02-05CH11231 using NERSC award FES-ERCAP0027638.}

\appendix
\section{Distinct properties of the $n=1$ mode {different from} $n=2$--$5$ modes}\label{app:a}

	The PFS, continuum, and 2D structure of the $n=1$ {mode show distinct features different from a typical TAE or EPM.} As shown in Figures~\ref{fig:pfs_n1_m1_5} and \ref{fig:pfs_n1_m6_9}, {unlike a TAE or EPM}, the $n=1$ mode does not exhibit any dominant neighbouring $m$-harmonic coupling; instead, the {$m=1,2,4,5,8$ and $9$ components all} have comparable amplitudes, forming a broad poloidal spectrum. Moreover, Figure~\ref{fig:continuum_n1} shows that its frequency lies well below the TAE gap. {Finally, the 2D ballooning structure of the $n=1$ mode is different from those of all other $n=2$--$5$ modes which are clearly anti-ballooning (Figure \ref{fig.con1}).}  Together, these features demonstrate that the $n=1$ branch possesses properties fundamentally different from the $n=2$--$5$ TAE-like EPMs. Thus the $n=1$ mode is excluded from the study on the parity transition of TAE/EPM in this work.
\begin{center}
	[Figure 11 about here.]\\
\end{center}

\section{{{Absence of shaping-induced parity transition in equilibria } with positive magnetic shear only}}\label{app:b}

To confirm the role of zero magnetic shear in the parity transition, a series of equilibria with {various elongations are} constructed, all sharing an analytically prescribed safety factor profile 
$q(\rho)=1.5+6.5\rho^{4}$ {with strong positive magnetic shear only}. {The $n=10$ mode is selected for analysis and demonstration due to its dominant growth rate among {all} toroidal harmonics considered.}

{Figures~\ref{fig:pos_k200},\ref{fig:pos_k150}, and \ref{fig:pos_k100} display the resulting mode structures for {decreasing} values of elongation, $\kappa = 2.00, 1.50,$ and $1.00$. In contrast to the weak shear case, all mode structures are identified as even-parity  and ballooning. Across the entire scan, the odd-parity, anti-ballooning mode is absent, and consequently, no parity transition is observed. }

\begin{center}
	[Figure 12 about here.]\\
\end{center}

{This confirms that the shaping-induced parity transition only takes place in {presence of zero magnetic} shear. This finding, {along with the observation that the} strong positive shear favors the sole existence of the even-parity ballooning mode, is consistent with {previous theory predictions in general} ~\cite{fuExistenceCoreLocalized1995,10.1063/1.871537,berkMoreCoreLocalized1995}.}
\makeatletter
\setcounter{figure}{0}
\renewcommand{\thefigure}{\arabic{figure}}
\makeatother
\clearpage

\begin{figure}[H]
	\centering
	\subfloat[\label{flux.fig}]{\includegraphics[width=0.45\textwidth,height=7.5cm]{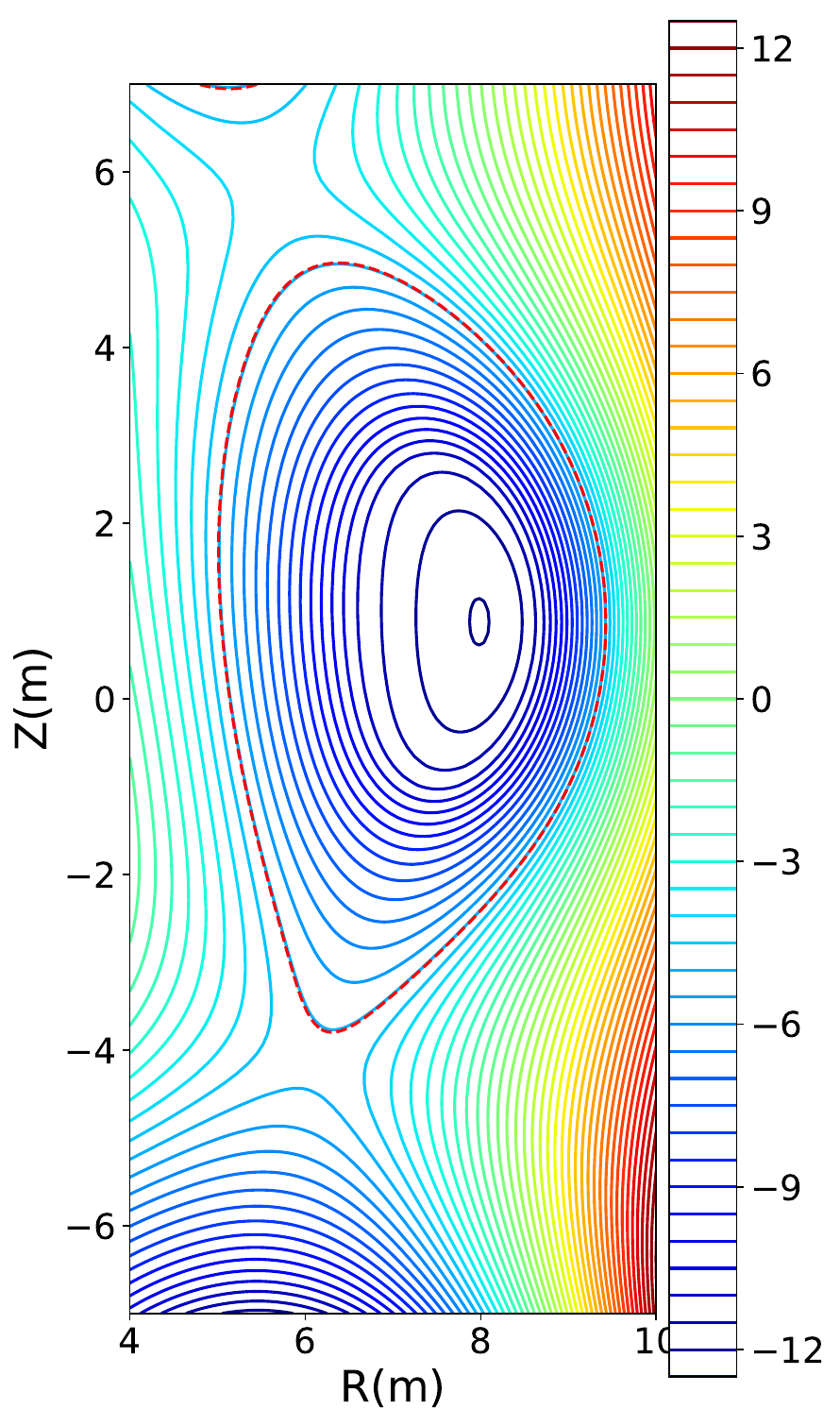}} 
	\vspace{0.2cm}
	\subfloat[\label{grid.fig}]{\includegraphics[width=0.4\textwidth,height=8cm]{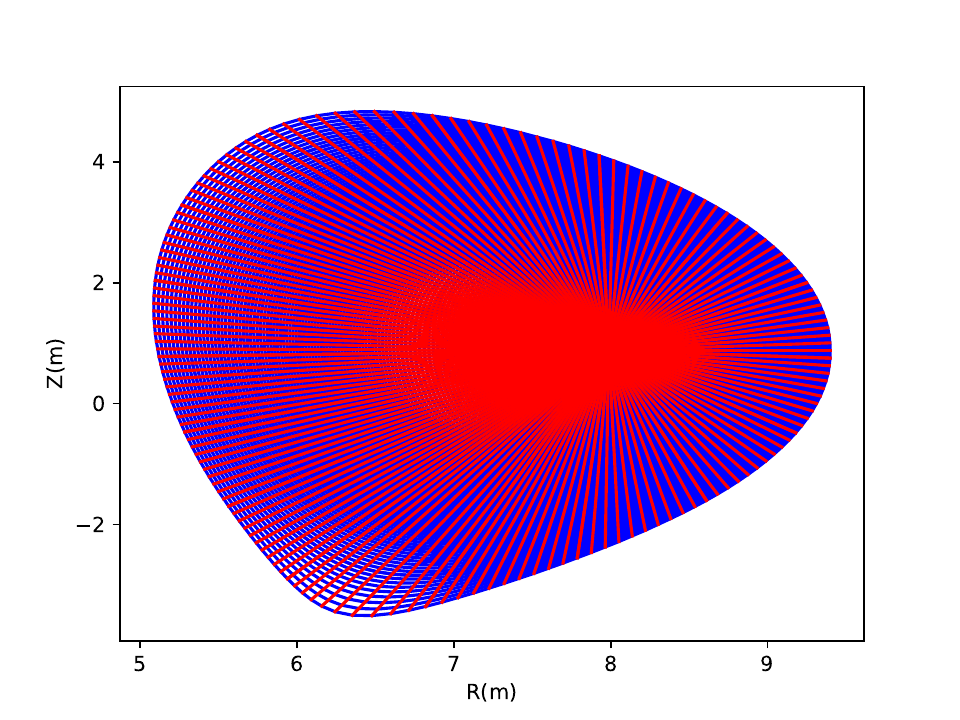}}\\
	\subfloat[\label{fig.safety}]{\includegraphics[width=0.45\textwidth,height=6.5cm]{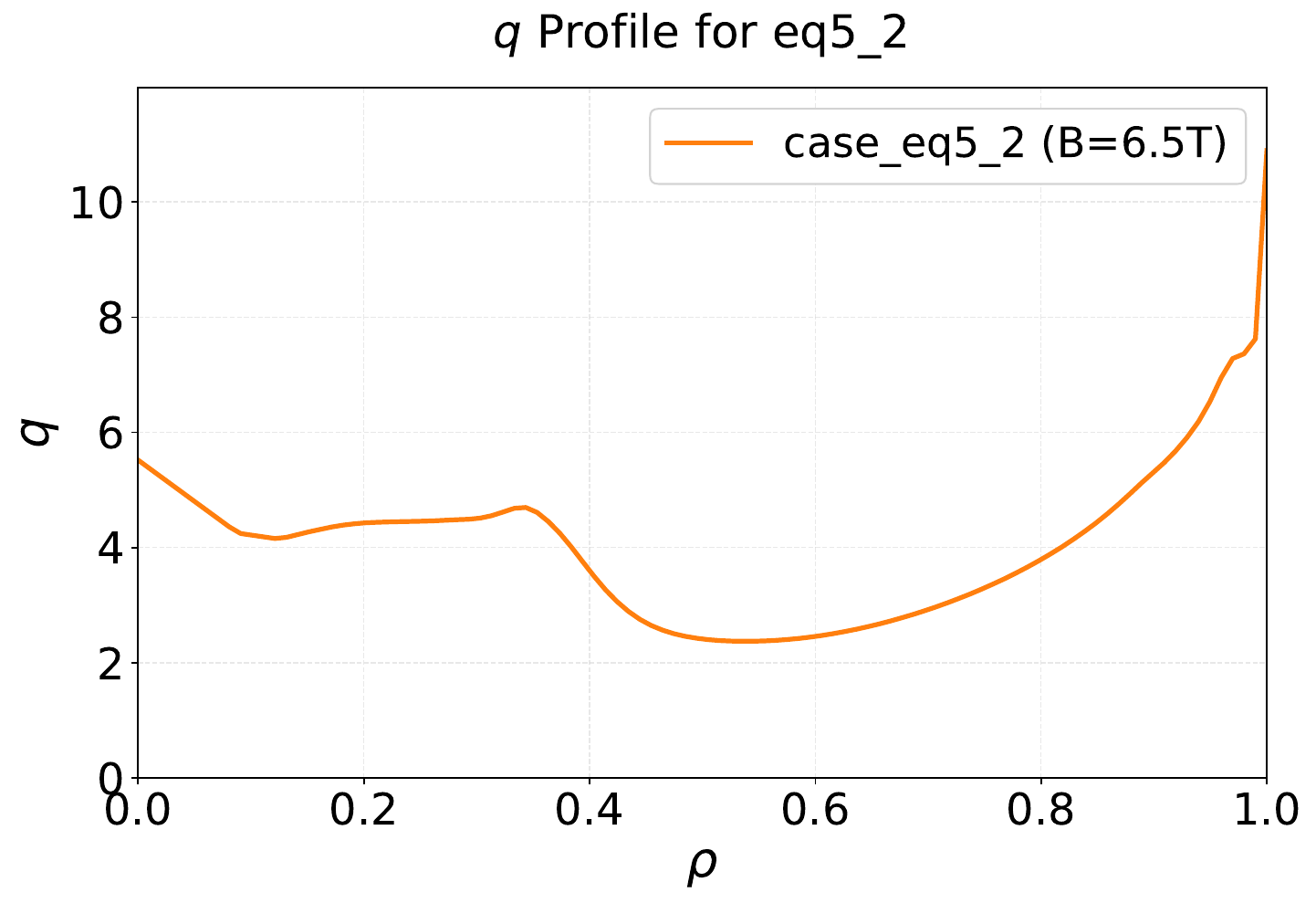}} 
	\vspace{1cm}
	\subfloat[\label{fig.pres}]{\includegraphics[width=0.45\textwidth,height=6.5cm]{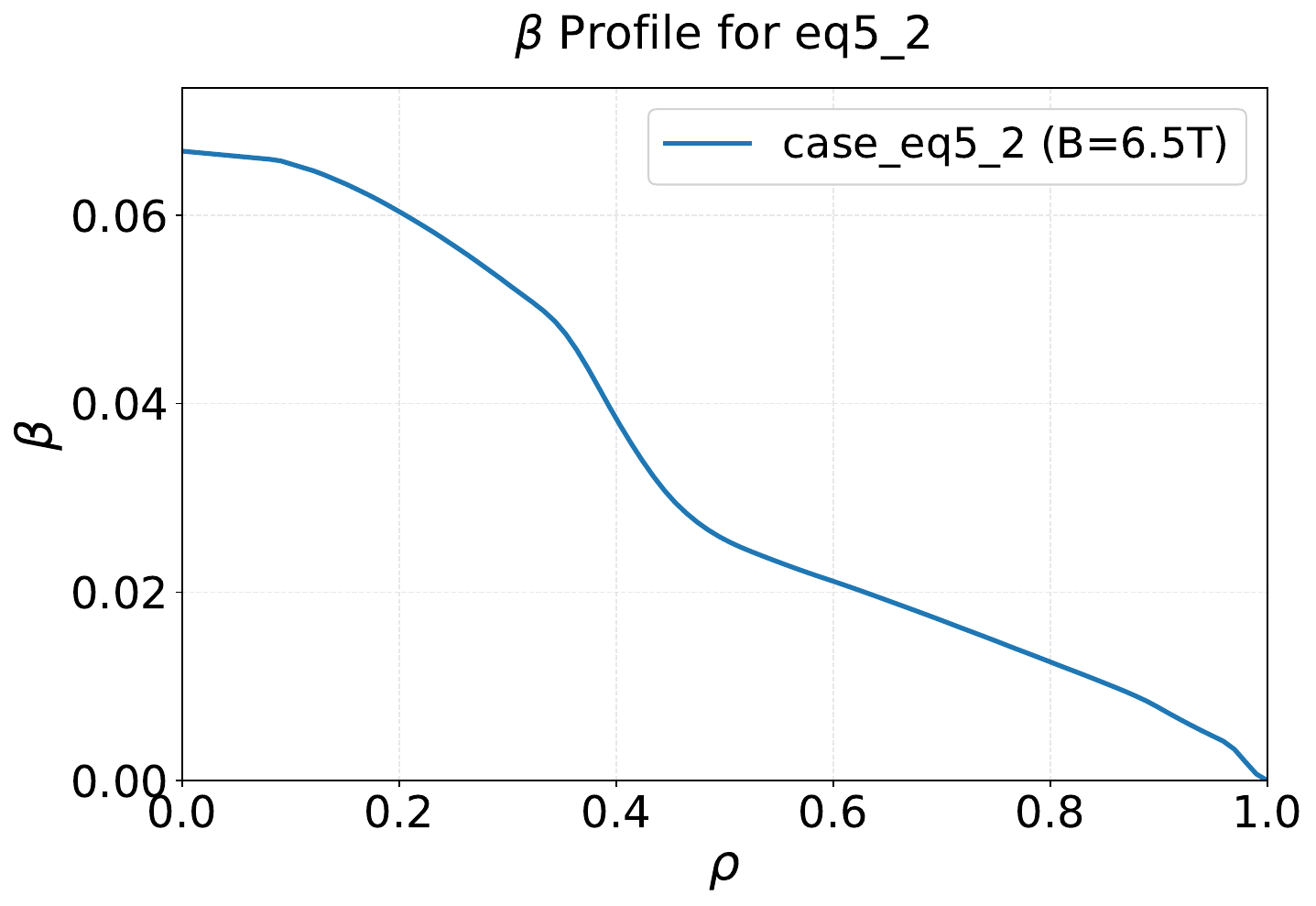}}
	\caption{(a) Contour plot of equilibrium poloidal flux in (R, Z) coordinate. The red curve represents the last closed flux surface (LCFS). (b) The mesh grid of flux coordinates used in the calculation. The blue lines represent the constant poloidal fluxes, and the red lines the poloidal angles. {(c) and (d) are the 1D radial profiles of safety factor and pressure. The radial coordinate $\rho$ represents the square root of the normalized poloidal flux, and the minimum value of the \(q\) profile is specified as \(q_{\text{min}} = 2.37\). The equilibrium is based on the CFETR case {eq5\_2}}}.
	\label{figure1}
\end{figure}
\clearpage

\begin{figure}[H]
	\centering
	\subfloat[\label{fig:freq}]{\includegraphics[width=0.8\textwidth,height=8cm]{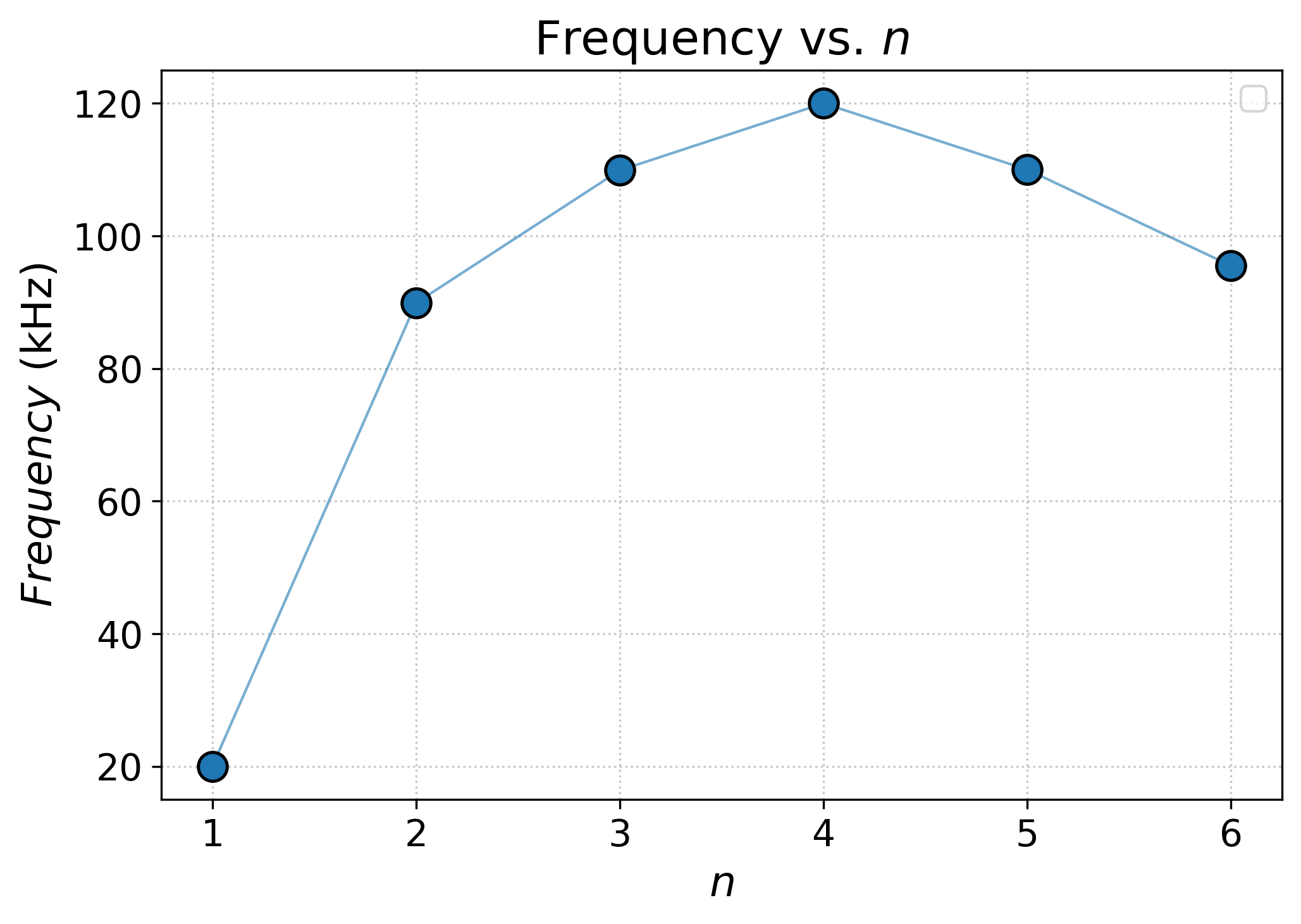}}
	\\
	\subfloat[\label{fig:growth}]{\includegraphics[width=0.8\textwidth,height=8cm]{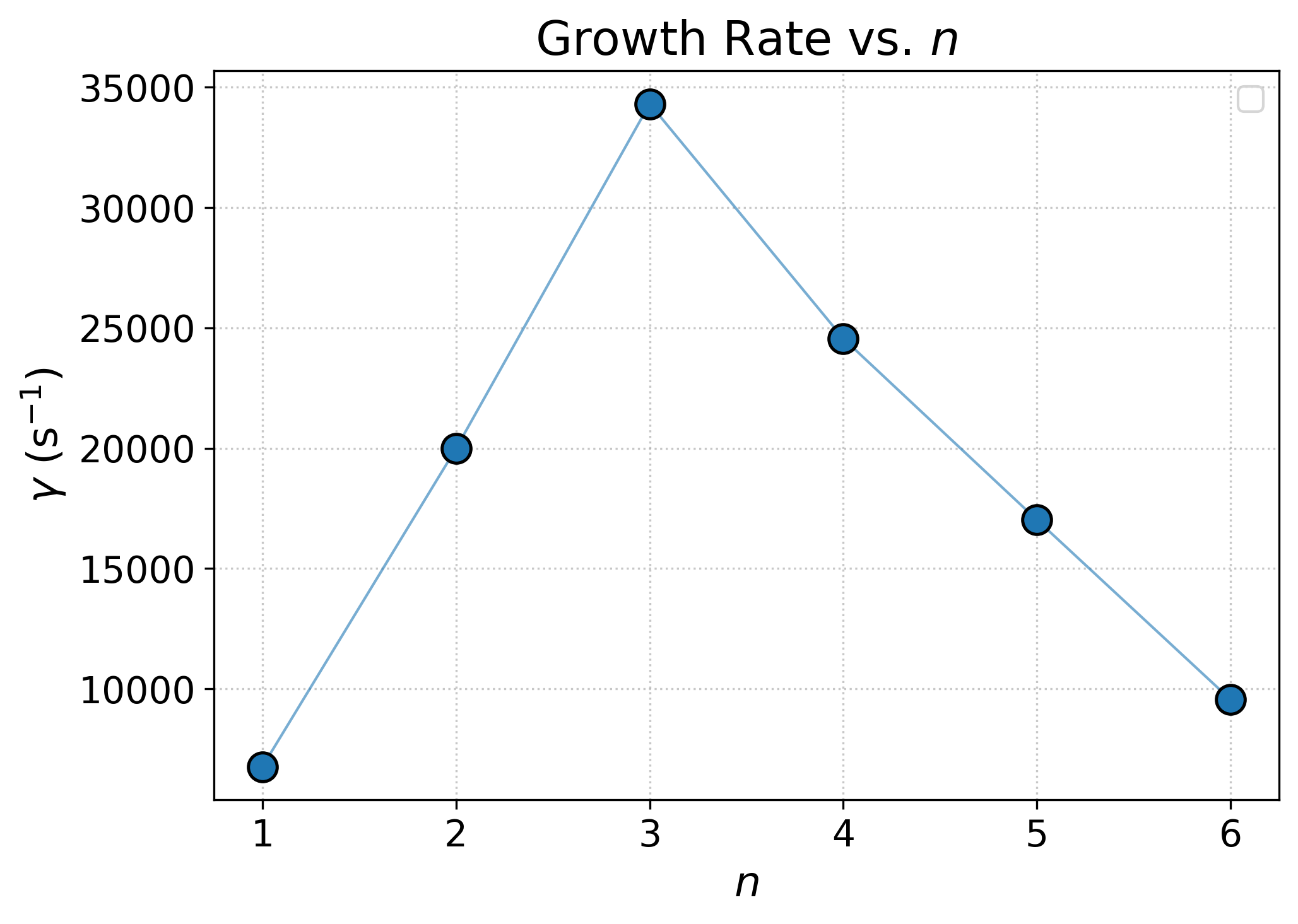}}
	\caption{The dependences of {the linear TAE/EPM} (a) frequency and (b) growth rate on the toroidal mode number.}
	\label{fig:fg}
\end{figure}
\clearpage

\begin{figure}[H]
	\centering
	\subfloat[]{\includegraphics[width=0.45\textwidth,height=6.2cm]{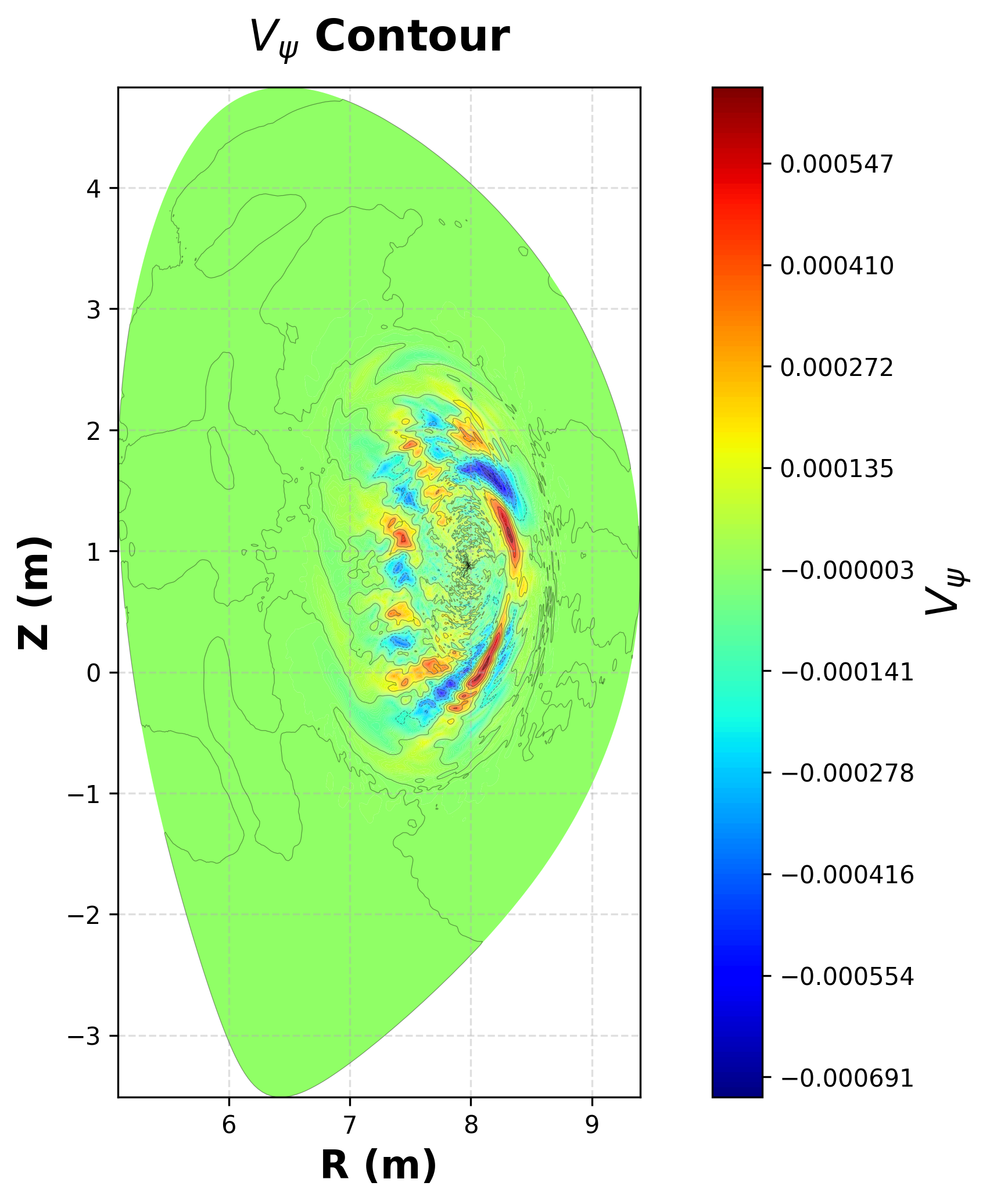}\label{fig.con1}}
	\subfloat[]{\includegraphics[width=0.45\textwidth,height=6.2cm]{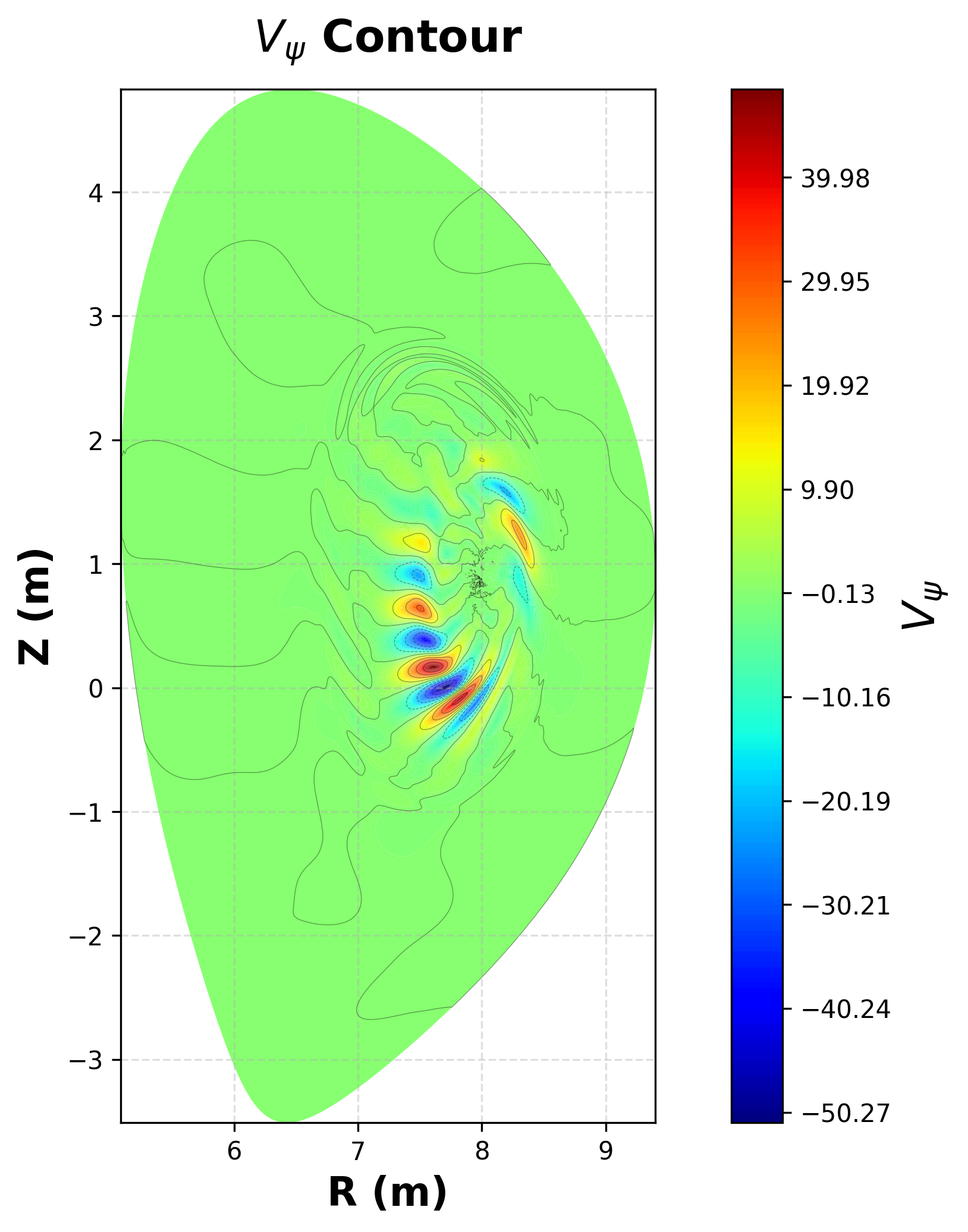}\label{fig.con2}}\\
	\subfloat[]{\includegraphics[width=0.45\textwidth,height=6.2cm]{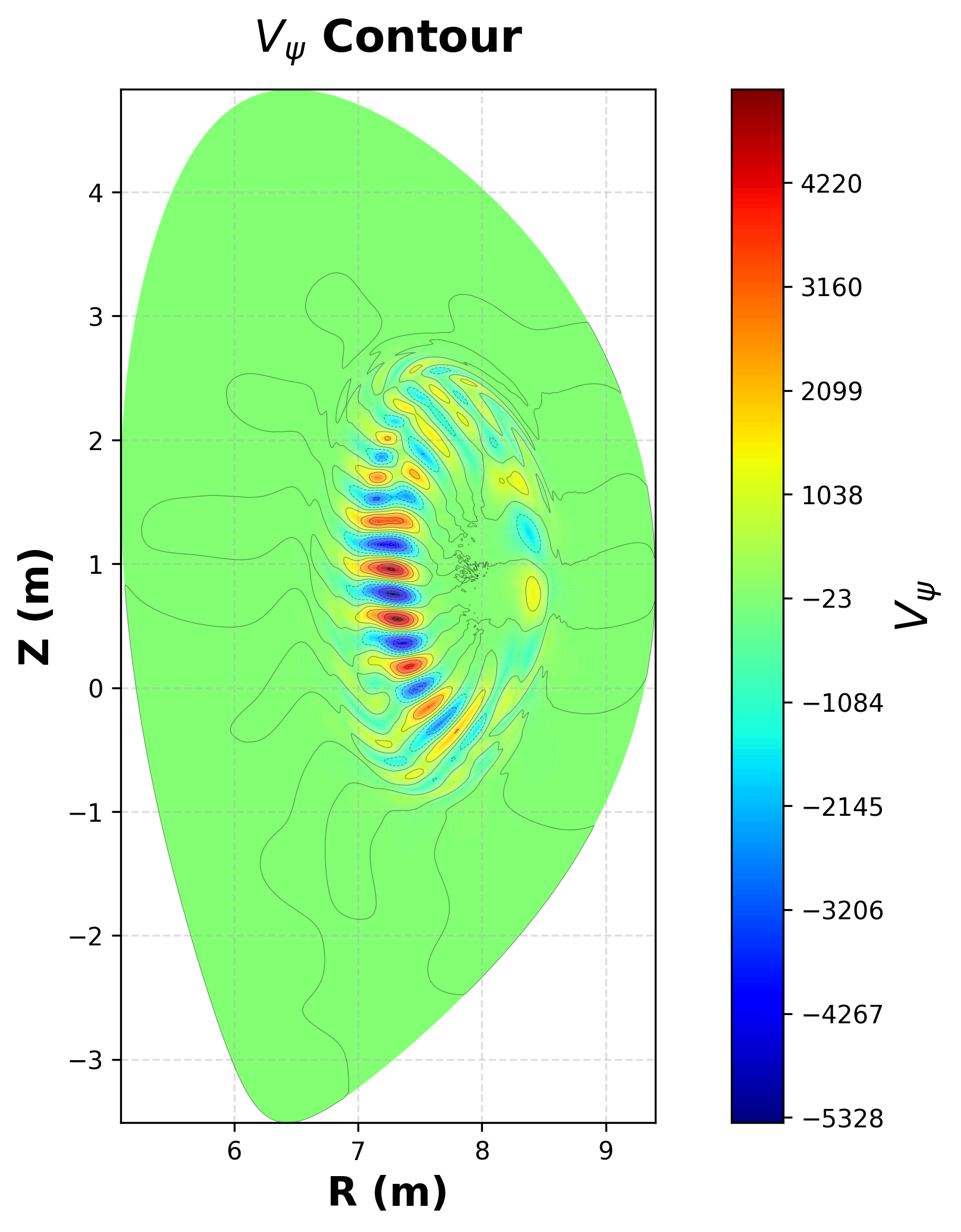}\label{fig.con3}}
	\subfloat[]{\includegraphics[width=0.45\textwidth,height=6.2cm]{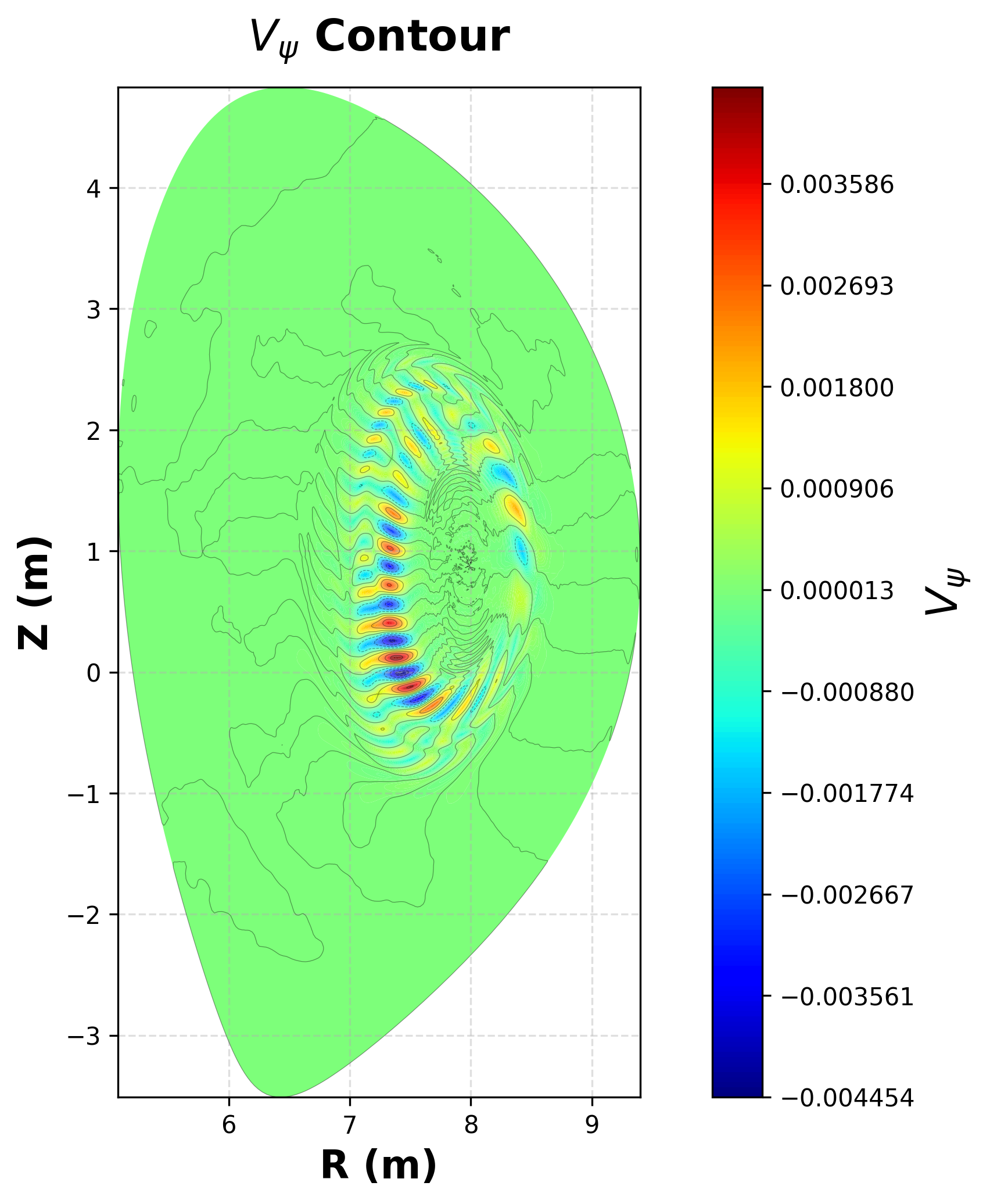}\label{fig.con4}}\\
	\subfloat[]{\includegraphics[width=0.45\textwidth,height=6.2cm]{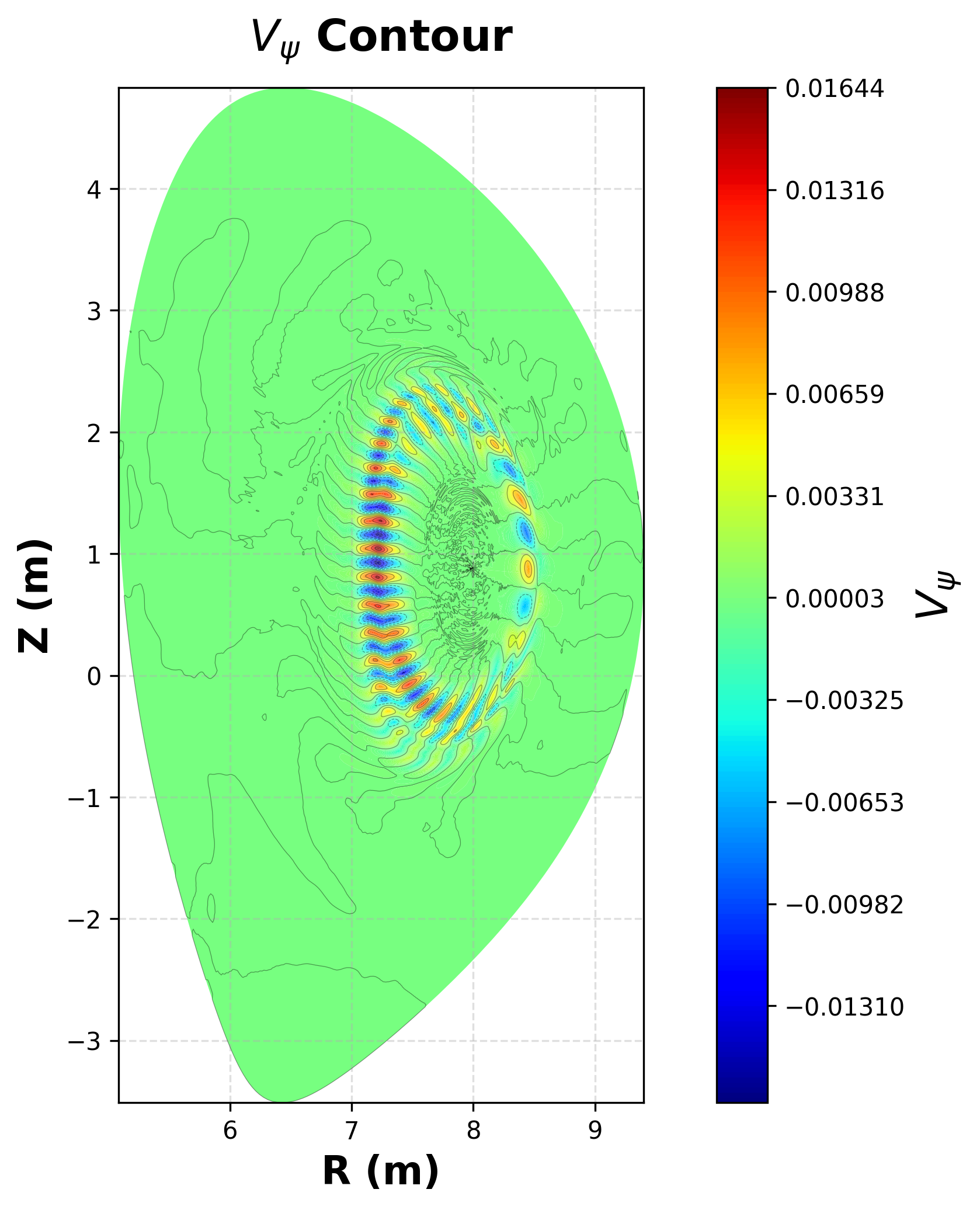}\label{fig.con5}}	
	\subfloat[]{\includegraphics[width=0.45\textwidth,height=6.2cm]{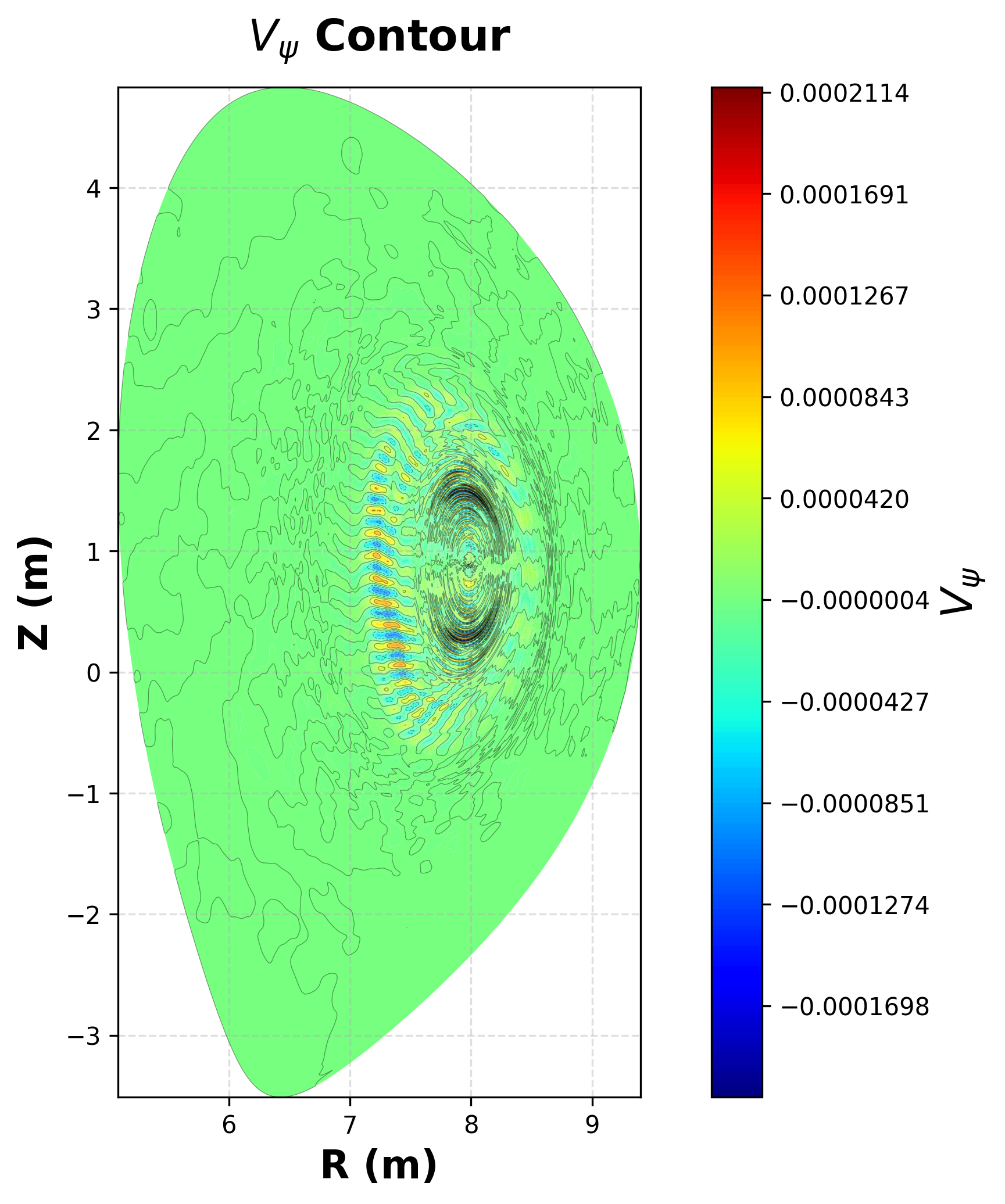}\label{fig.con6}}	
	\caption{\label{fig:contour} {Contours of perturbed normal velocity component from NIMROD simulations for toroidal mode number}: (a) n=1, (b) n=2, (c) n=3, (d) n=4, (e) n=5 and (f) n=6.}
\end{figure}
\clearpage

\begin{figure}[H]
	\centering

	\subfloat[$n=3$ PFS \label{fig:pfs_n3_org}]{
		\includegraphics[width=0.68\textwidth,height=7.5cm]{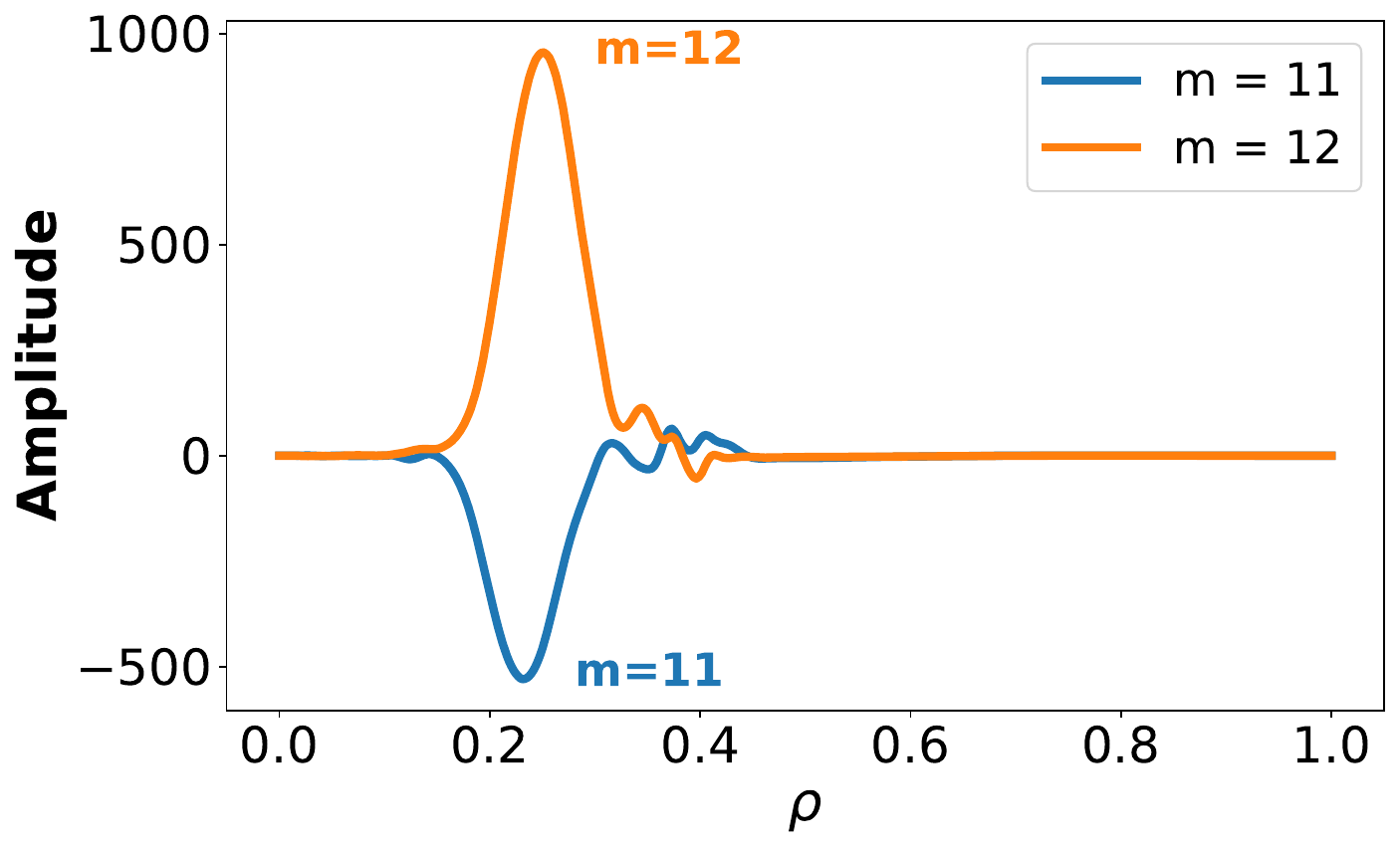}
	}\hfill
	
	\subfloat[$n=3$ Alfv\'{e}n continuum \label{fig:continuum_n3}]{
		\includegraphics[width=0.75\textwidth,height=7.5cm]{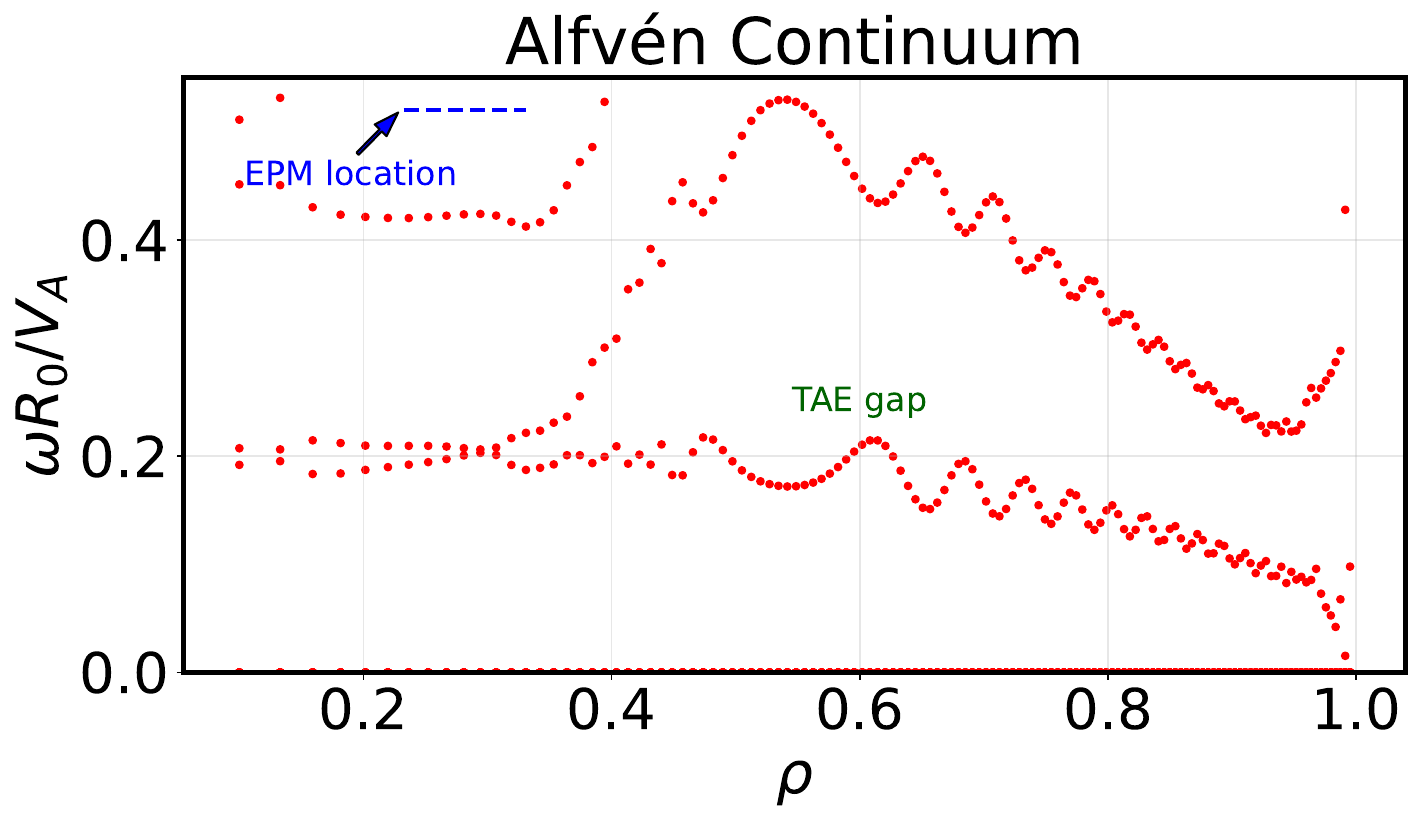}
	}
	
	\caption{
		{
			(a) Radial profiles of {two most dominant} poloidal Fourier components with toroidal mode number $n=3$ from NIMROD simulation, and  
			(b) the corresponding Alfv\'{e}n continuum calculated using GTAW.
		}
	}
	\label{fig:combined_pfs_continuum}
\end{figure}
\clearpage

\begin{figure}[H]
	\centering
	
	\subfloat[Frequency vs $q_{\text{min}}$ \label{fig:freq_q}]{
		\includegraphics[width=0.45\textwidth,height=6.1cm]{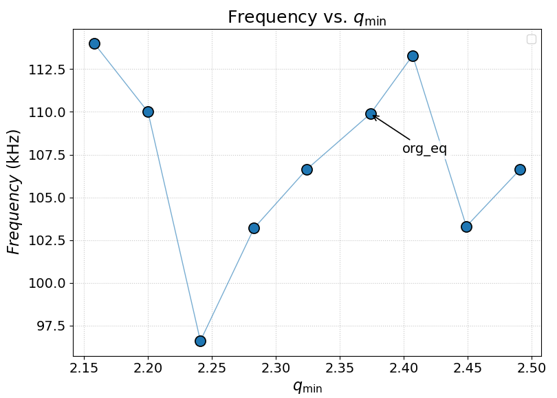}
	}\hfill
	\subfloat[Growth rate vs $q_{\text{min}}$ \label{fig:growth_q}]{
		\includegraphics[width=0.45\textwidth,height=6.1cm]{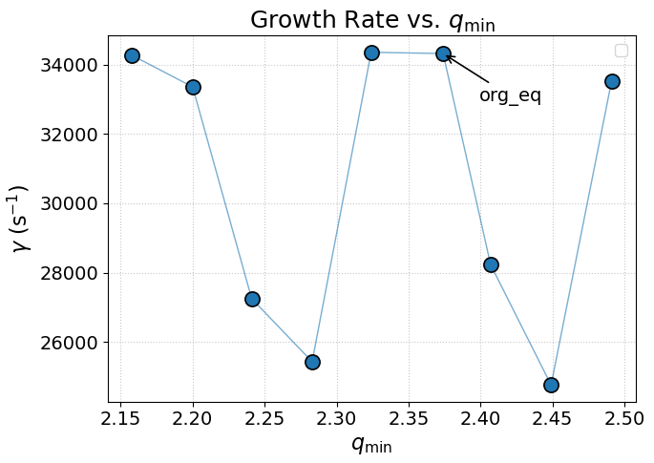}
	}\\[0.5em]
	\subfloat[$q_{\text{min}}=2.15$ \label{fig:qmin215}]{
		\includegraphics[width=0.45\textwidth,height=6.5cm]{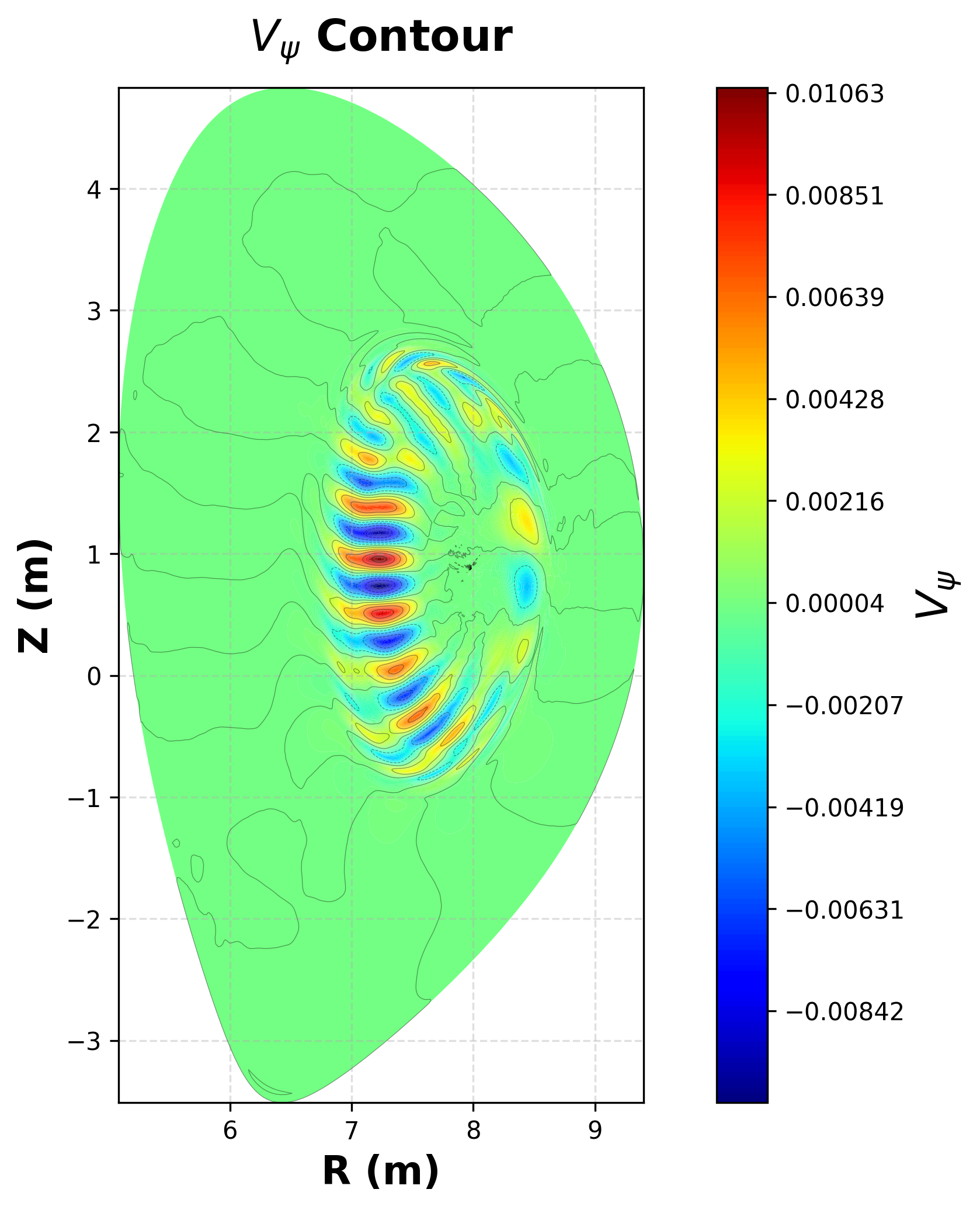}
	}\hfill
	\subfloat[$q_{\text{min}}=2.50$ \label{fig:qmin250}]{
		\includegraphics[width=0.45\textwidth,height=6.5cm]{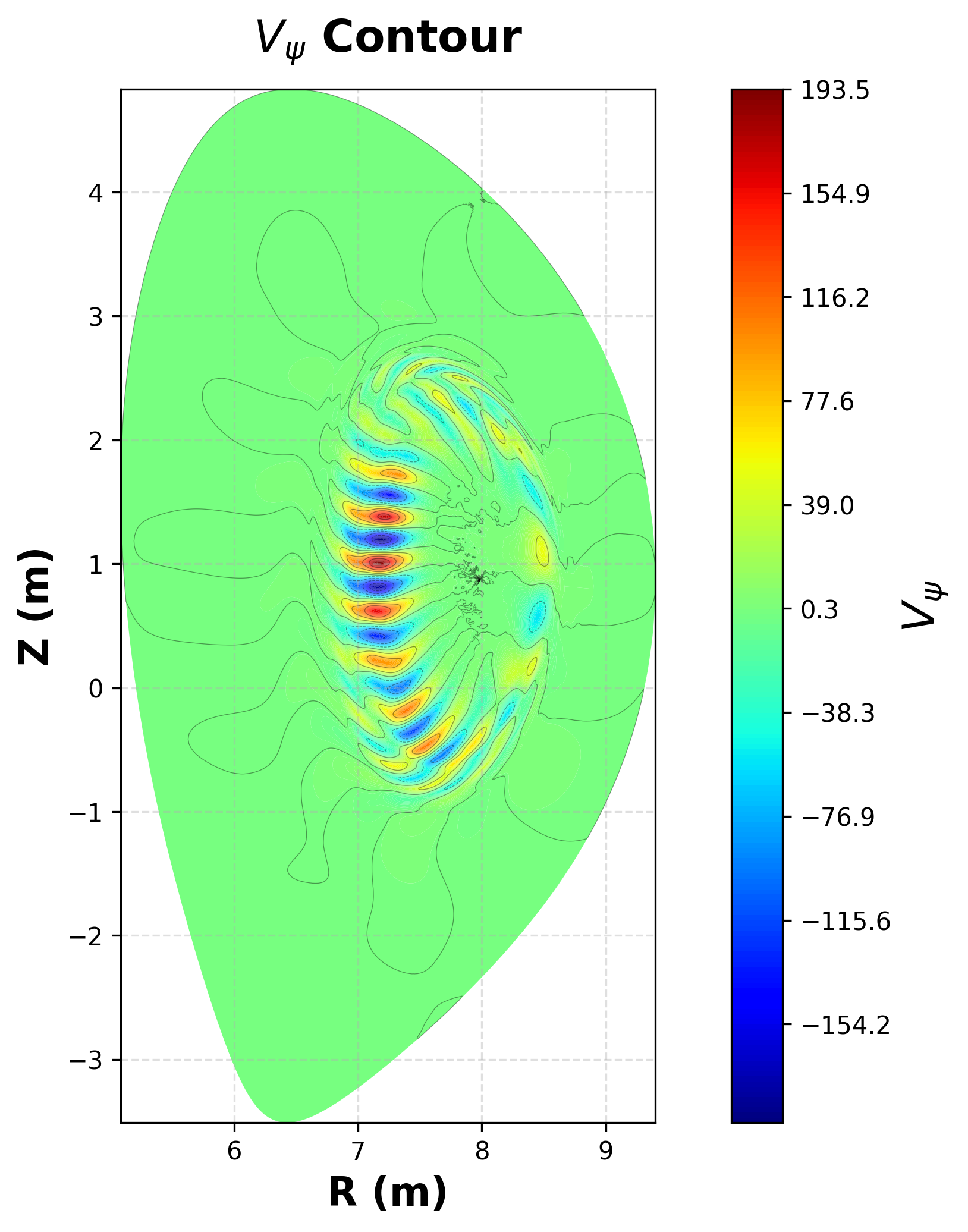}
	}
	
	\caption{{The dependences of (a) frequency and (b) growth rate on minimum safety factor $q_{\text{min}}$ for the toroidal mode number $n=3$, 
			and contours of perturbed normal velocity component from NIMROD simulations for 
			(c) $q_{\text{min}}=2.15$ and (d) $q_{\text{min}}=2.50$ modes.}
	}
	\label{fig:combined_qmin}
\end{figure}
\clearpage

\begin{figure}[H]
	\centering
	\subfloat[Frequency vs $\beta_h$ \label{fig:freq_beta}]{
		\includegraphics[width=0.45\textwidth,height=6cm]{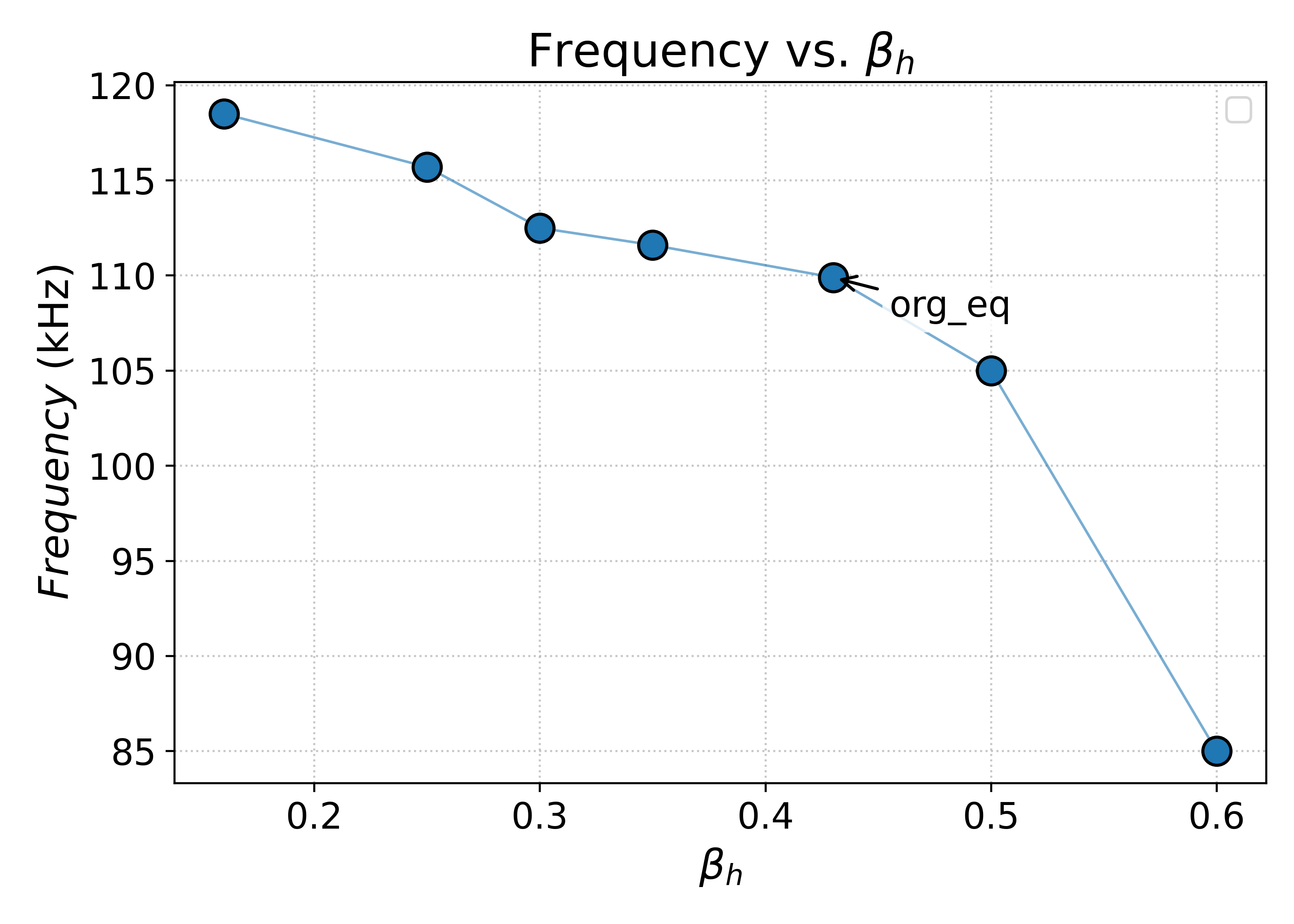}
	}\hfill
	\subfloat[Growth rate vs $\beta_h$ \label{fig:growth_beta}]{
		\includegraphics[width=0.45\textwidth,height=6cm]{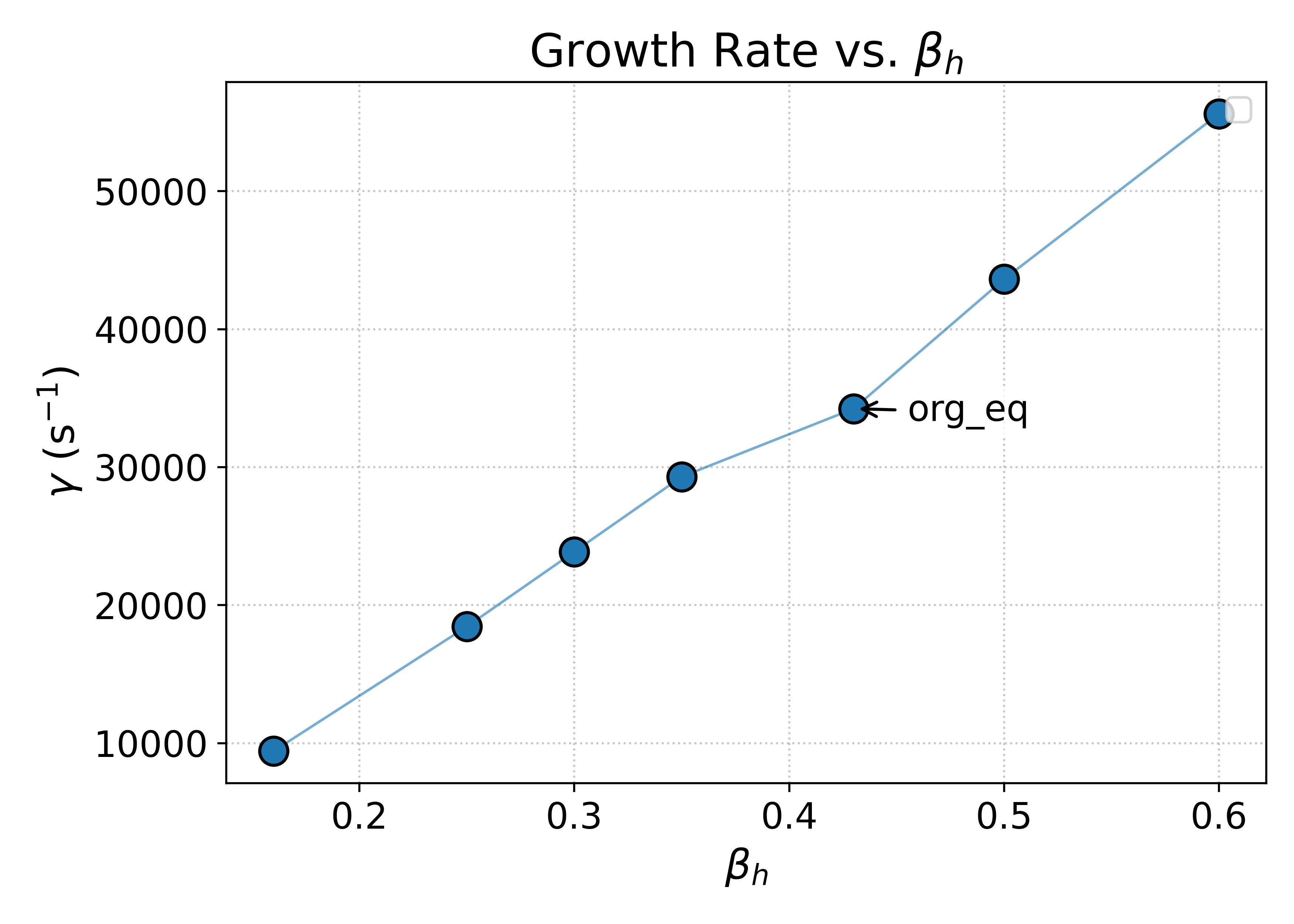}
	}\\[0.5em]
	
	\subfloat[$\beta_h = 0.16$ \label{fig:beta016}]{
		\includegraphics[width=0.45\textwidth,height=6cm]{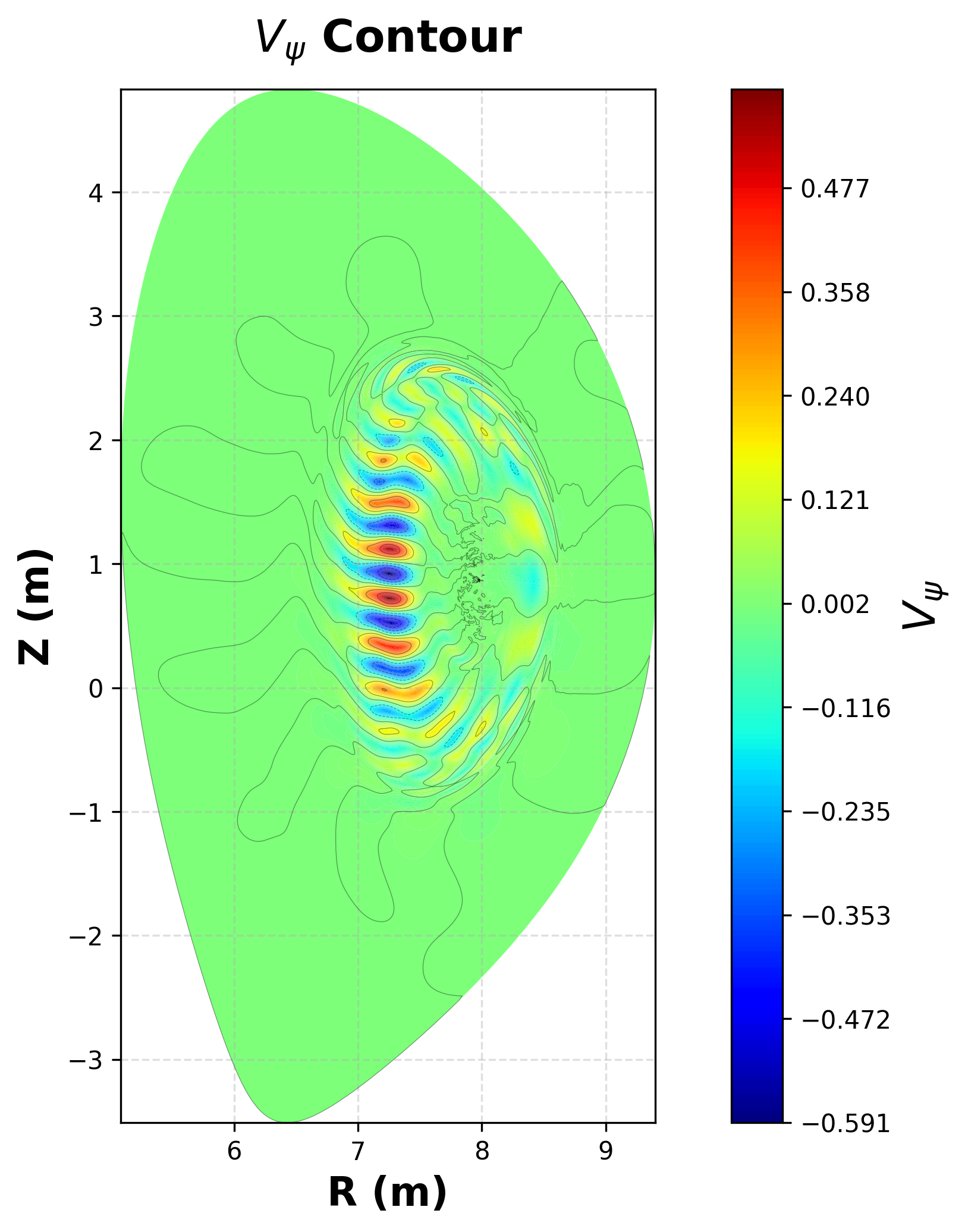}
	}\hfill
	\subfloat[$\beta_h = 0.60$ \label{fig:beta060}]{
		\includegraphics[width=0.45\textwidth,height=6cm]{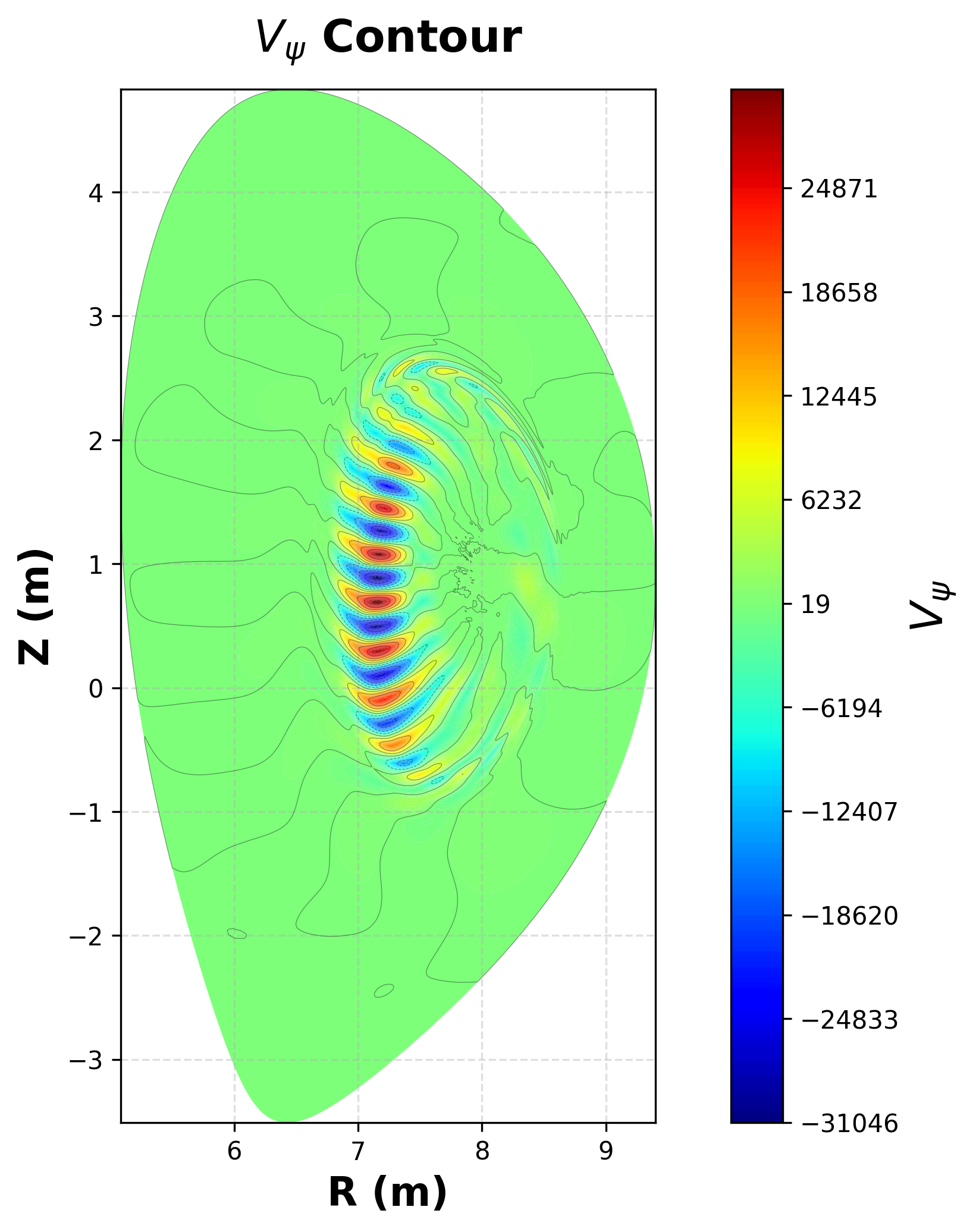}
	}
	
	\caption{{The dependences of (a) frequency and (b) growth rate on energetic particle $\beta$ fraction $\beta_h$ for the toroidal mode number $n=3$, and contours of perturbed normal velocity component from NIMROD simulations for (c) $\beta_h = 0.16$ and (d) $\beta_h = 0.60$ modes.}
	}
	\label{fig:combined_beta}
\end{figure}
\clearpage

\begin{figure}[H]
	\centering
	
	\subfloat[$\beta$ profiles \label{pres2}]{
		\includegraphics[width=0.47\textwidth,height=6.2cm]{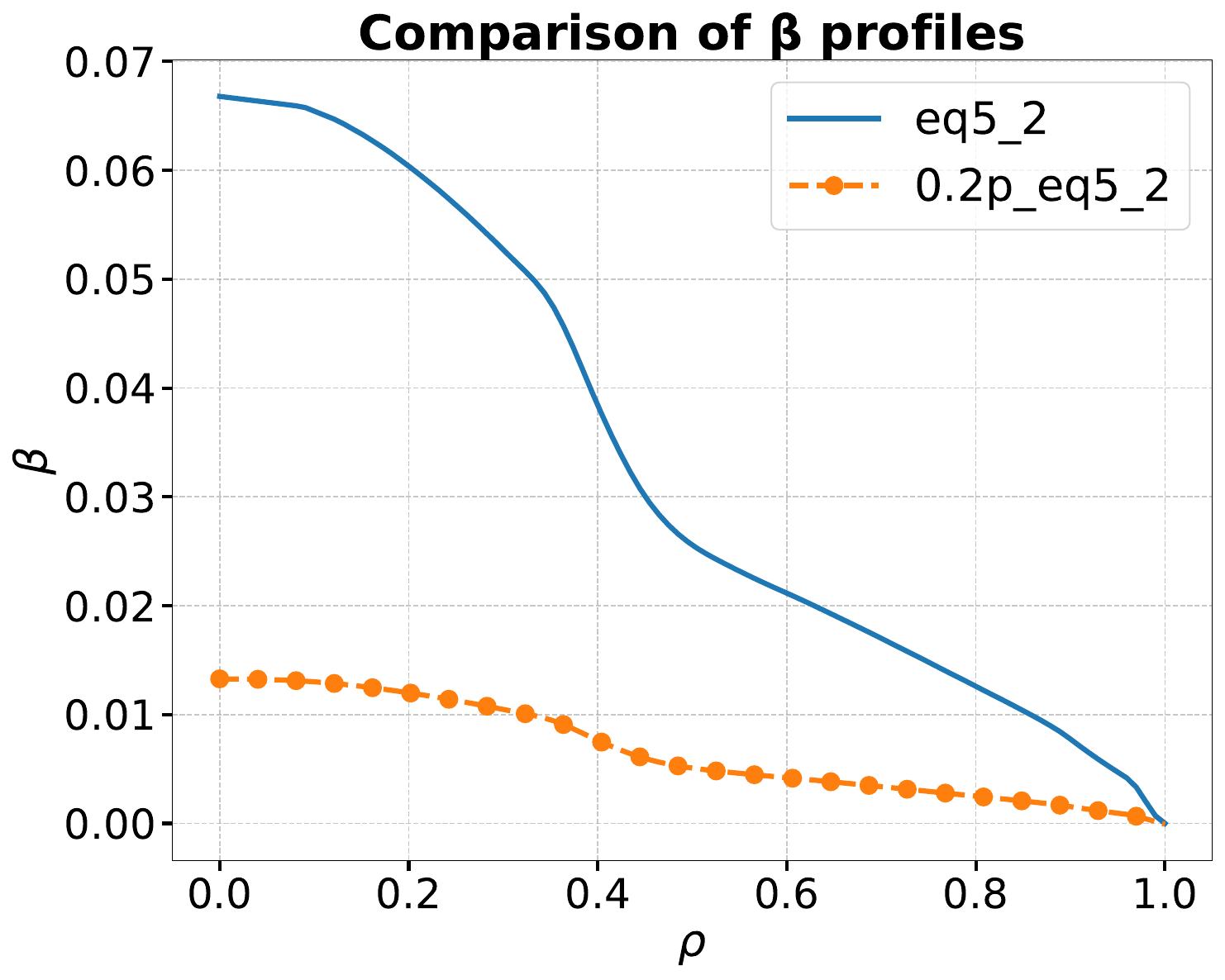}
	}\hfill
	\subfloat[Safety factor profiles \label{safe2}]{
		\includegraphics[width=0.47\textwidth,height=6.2cm]{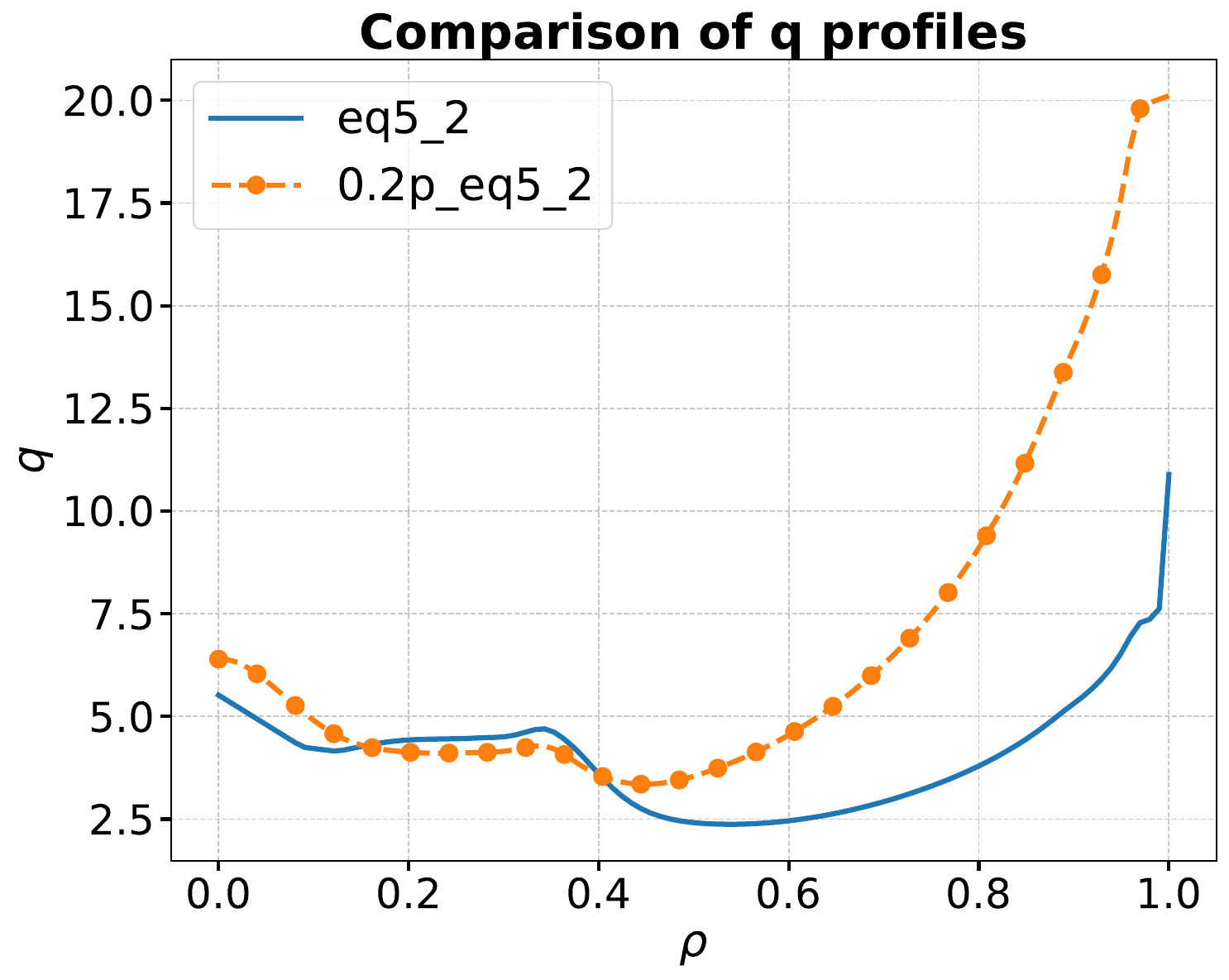}
	}\\[0.8em]

	\subfloat[Alfv\'en continuum of the $n=3$ mode for the modified equilibrium
	\label{continuum_new_n3}]{
		\includegraphics[width=0.75\textwidth,height=7.5cm]{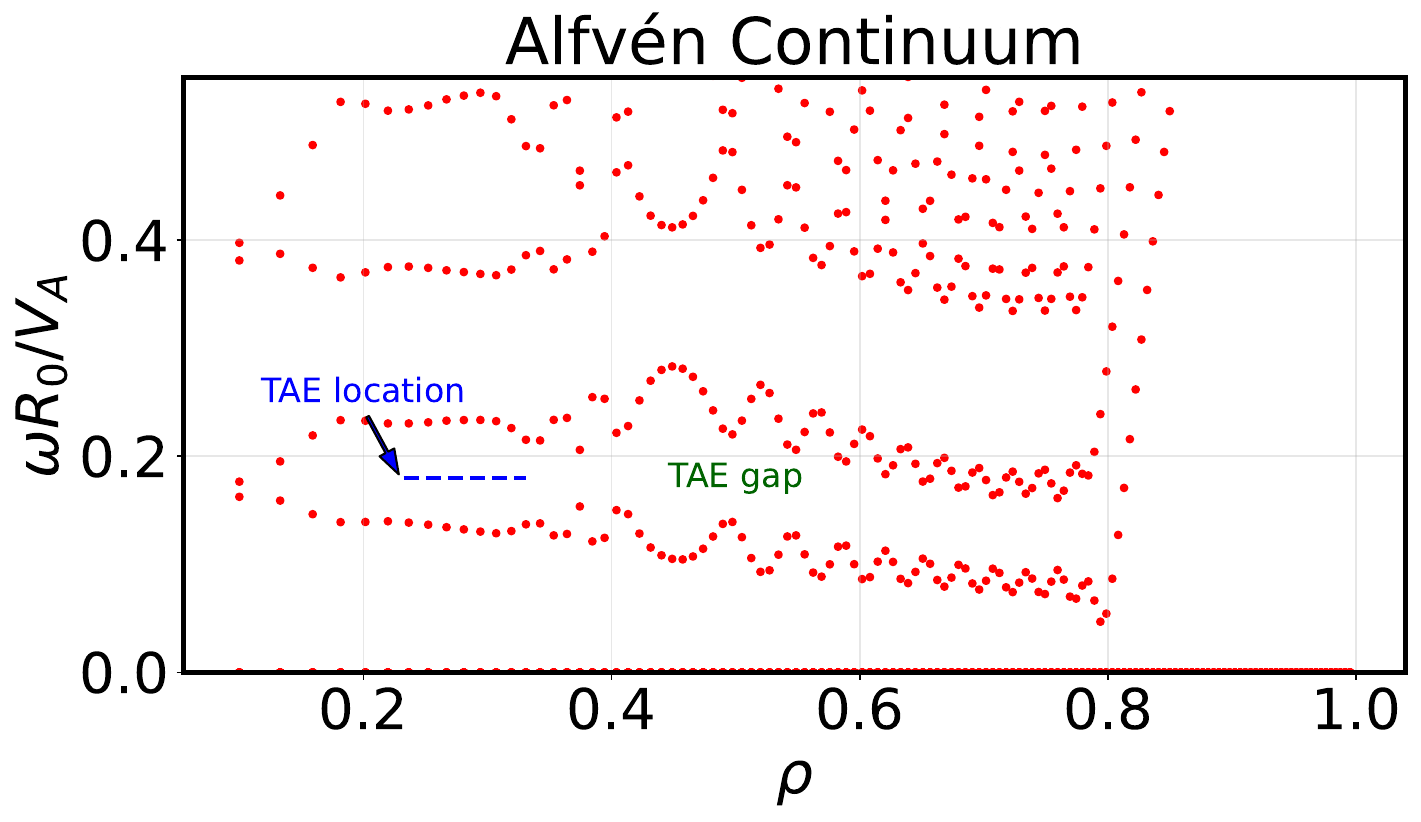}
	}

	\caption{
		(a) $\beta$ and (b) safety factor profiles 
		{of the CFETR case equilibrium eq5\_2 and the case with reduced plasma $\beta$}, 
		and (c) the corresponding Alfv\'en continuum of $n=3$ mode 
		for the equilibrium with reduced $\beta$.
	}
	\label{fig:combined_eq_new}
\end{figure}
\clearpage

\begin{figure}[H]
	\centering
	
	\subfloat[]{
		\includegraphics[width=0.65\textwidth,height=6.8cm]{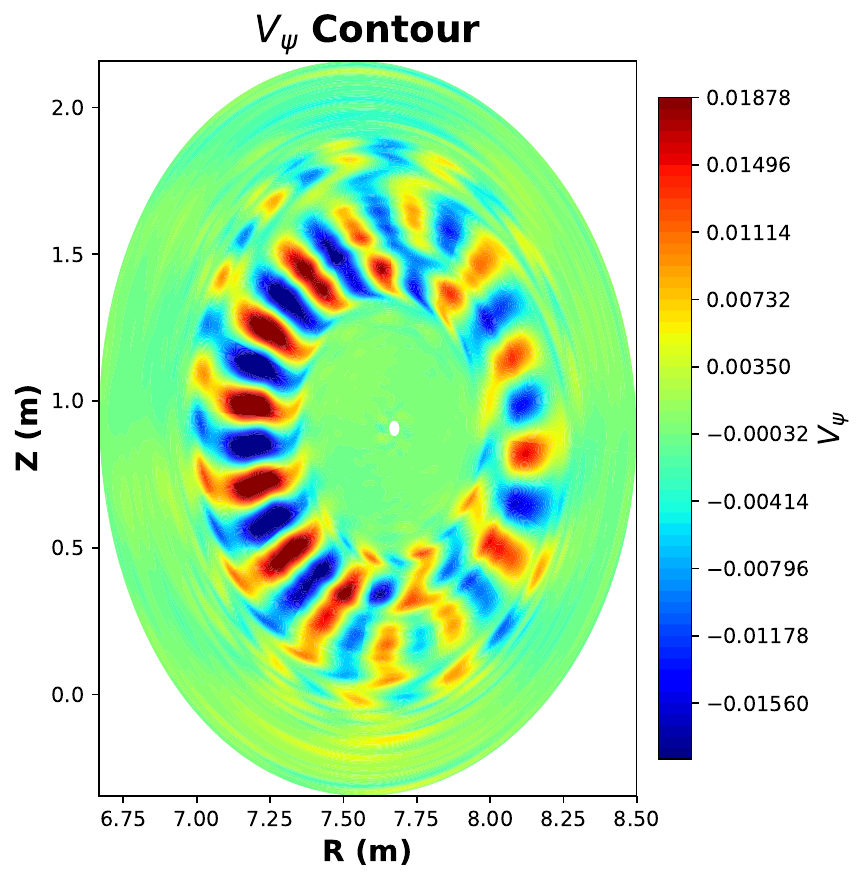}
		\label{con_new_n3}
	}\\[0.8em]
	
	\subfloat[]{
		\includegraphics[width=0.65\textwidth,height=6.8cm]{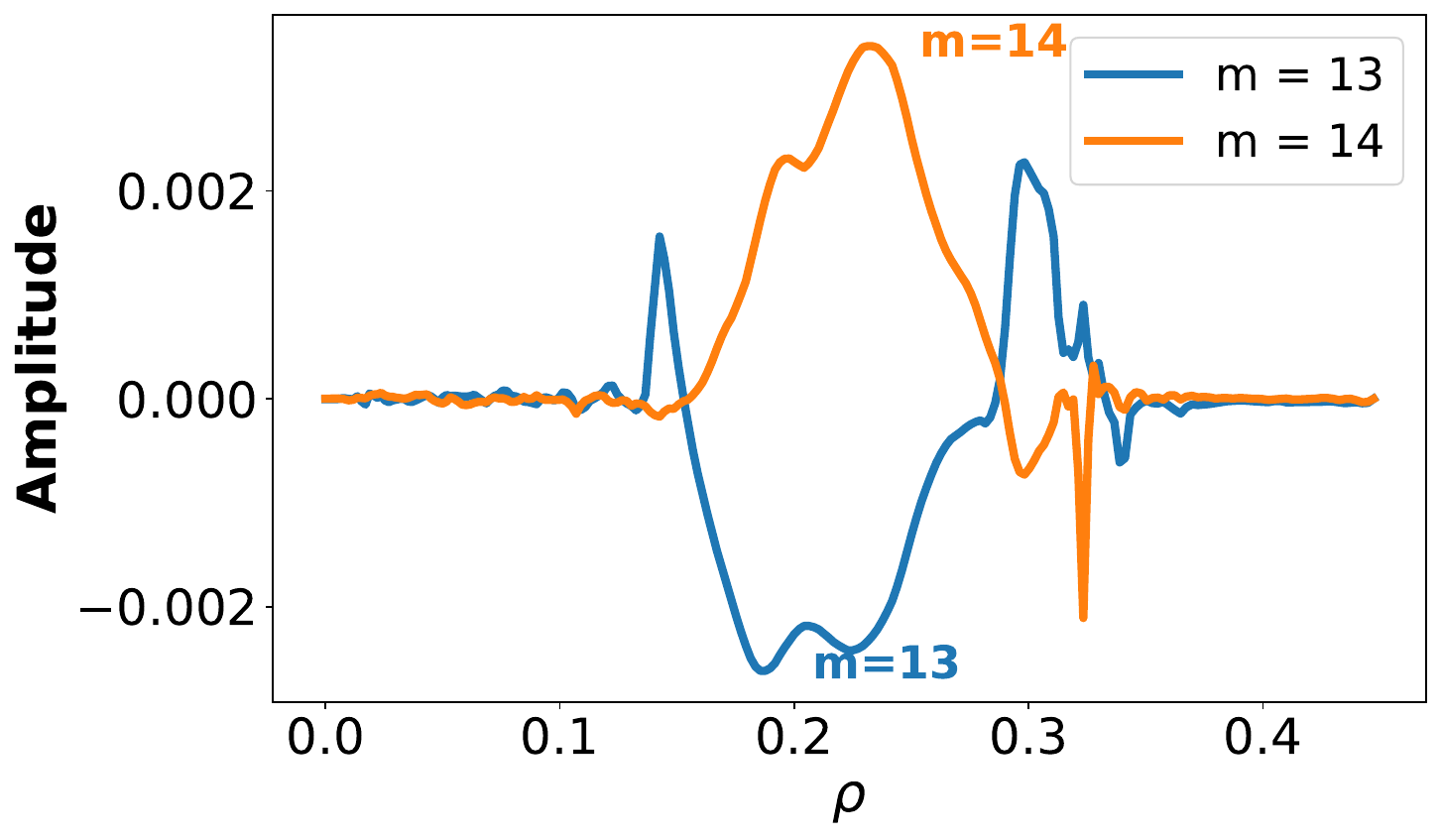}
		\label{pfs_new_n3}
	}
	
	\caption{
		{
			(a) Contour of the perturbed normal velocity component, and 
			(b) radial profiles {two most dominant} of poloidal Fourier components {from NIMROD simulations}, both corresponding to the $n=3$ mode 
			in the equilibrium {with reduced $\beta$}.}
	}
	\label{fig:combined_vpsi_pfs}
\end{figure}
\clearpage

\begin{figure}[H]
	\centering
	\includegraphics[width=0.6\textwidth,height=8.5cm]{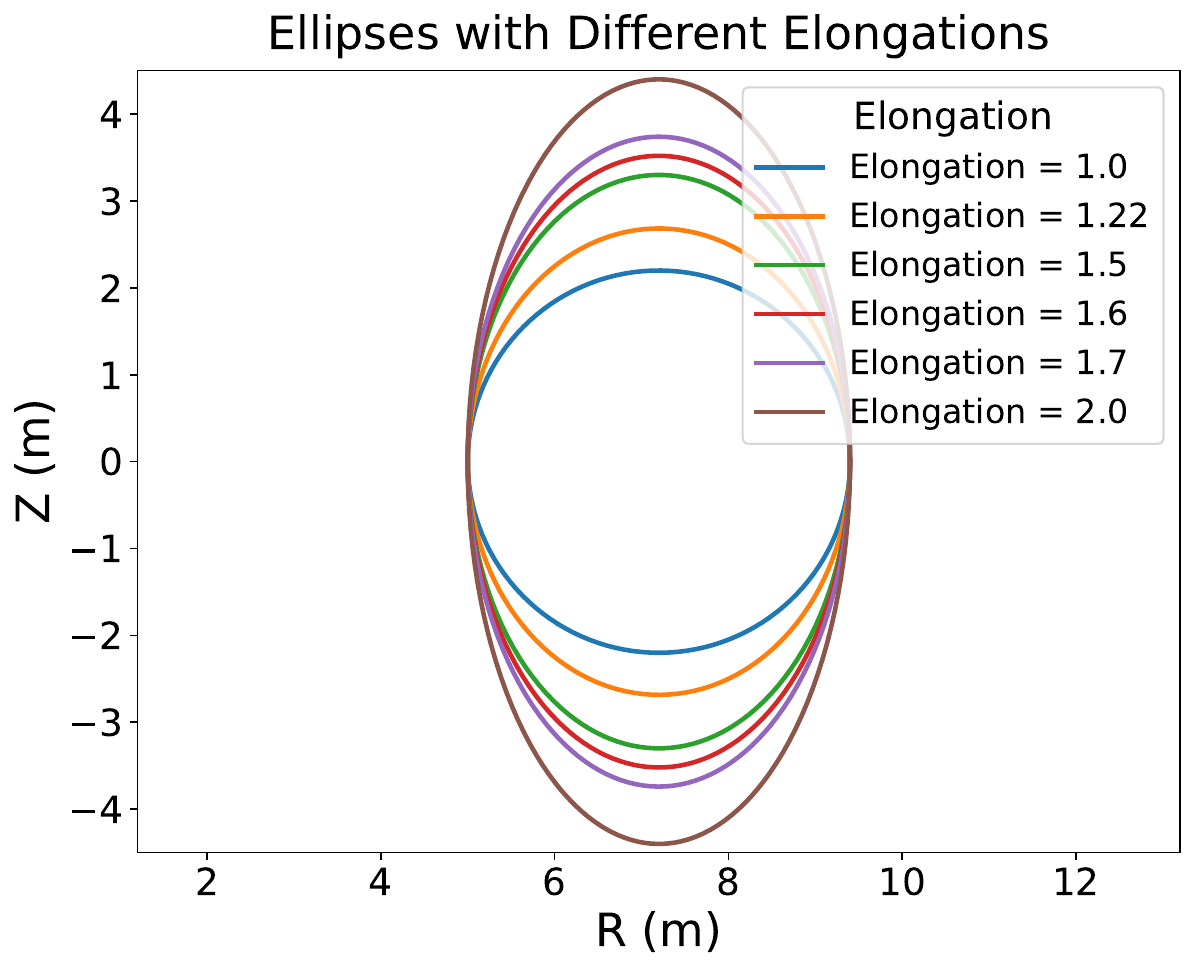}
	\caption{Last closed flux surfaces for equilibria with {various} elongations, ranging from $b/a=2.0$ to $b/a=1.0$, generated {using} the CHEASE code.}
	\label{ellipse}
\end{figure}
\clearpage

\begin{figure}[H]
	\centering
	
	\subfloat[$\kappa=2.0$ contour \label{con_ba20}]{
		\includegraphics[width=0.435\textwidth,height=5.2cm]{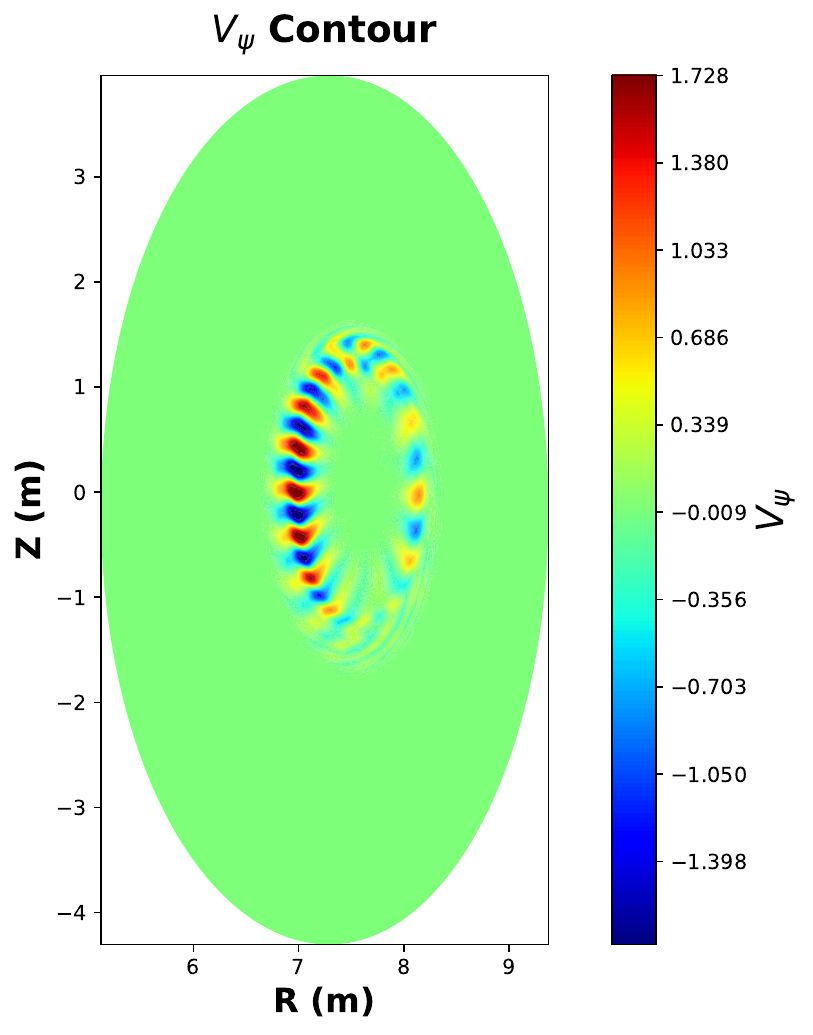}
	}\hfill
	\subfloat[$\kappa=2.0$ PFS \label{pfsba20}]{
		\includegraphics[width=0.435\textwidth,height=5.2cm]{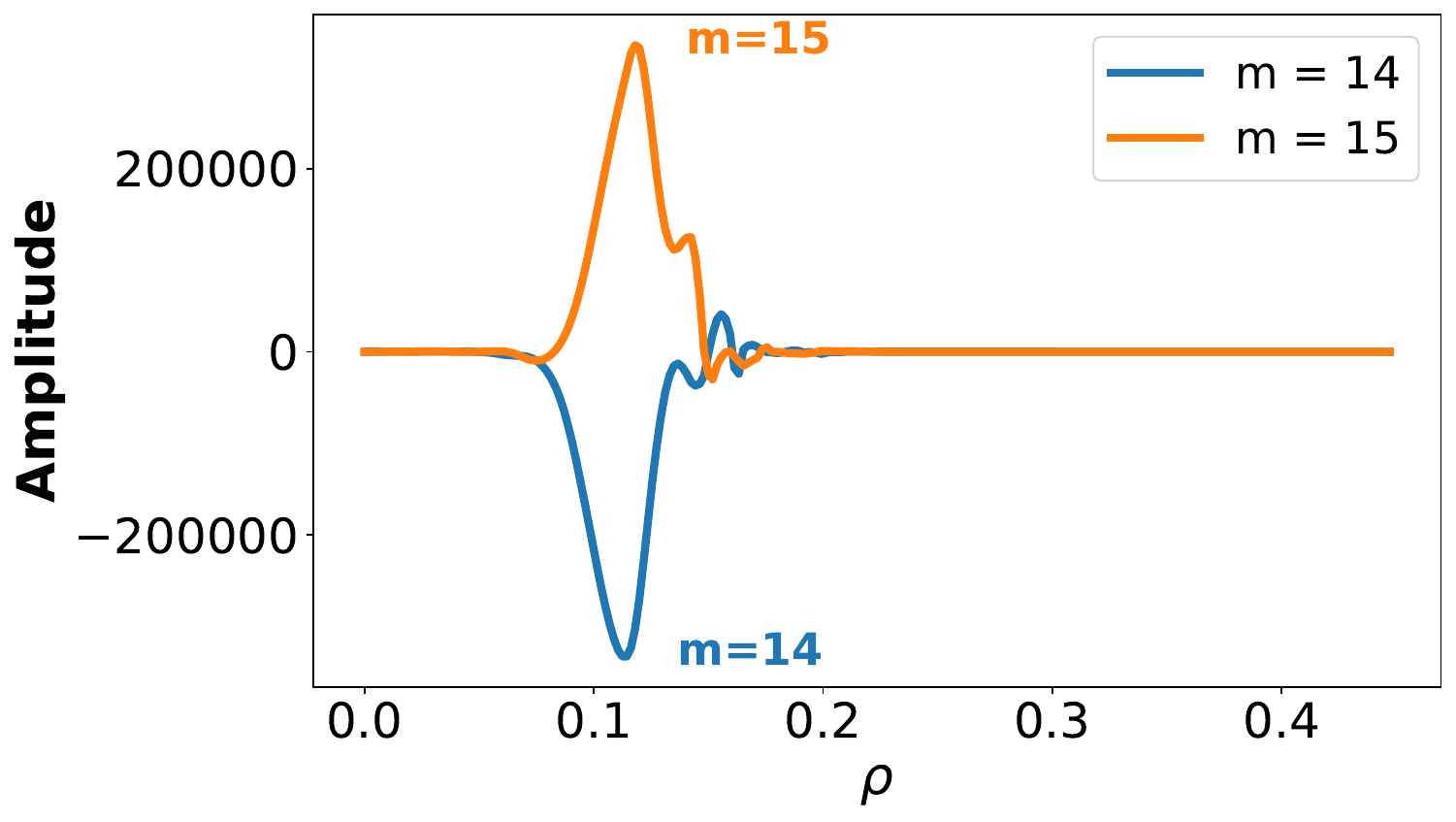}
	}\\[0.6em]

	\subfloat[$\kappa=1.6$ contour \label{con_ba16}]{
		\includegraphics[width=0.42\textwidth,height=5.2cm]{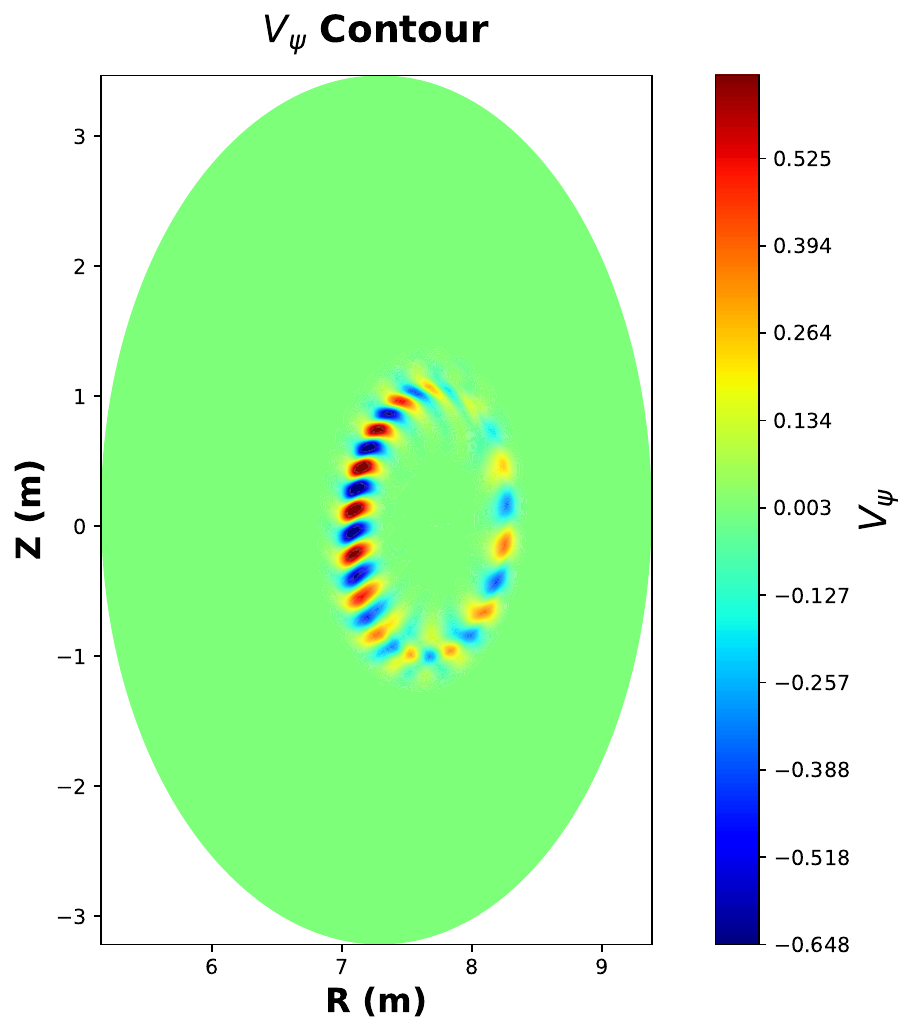}
	}\hfill
	\subfloat[$\kappa=1.6$ PFS \label{pfsba16}]{
		\includegraphics[width=0.45\textwidth,height=5.2cm]{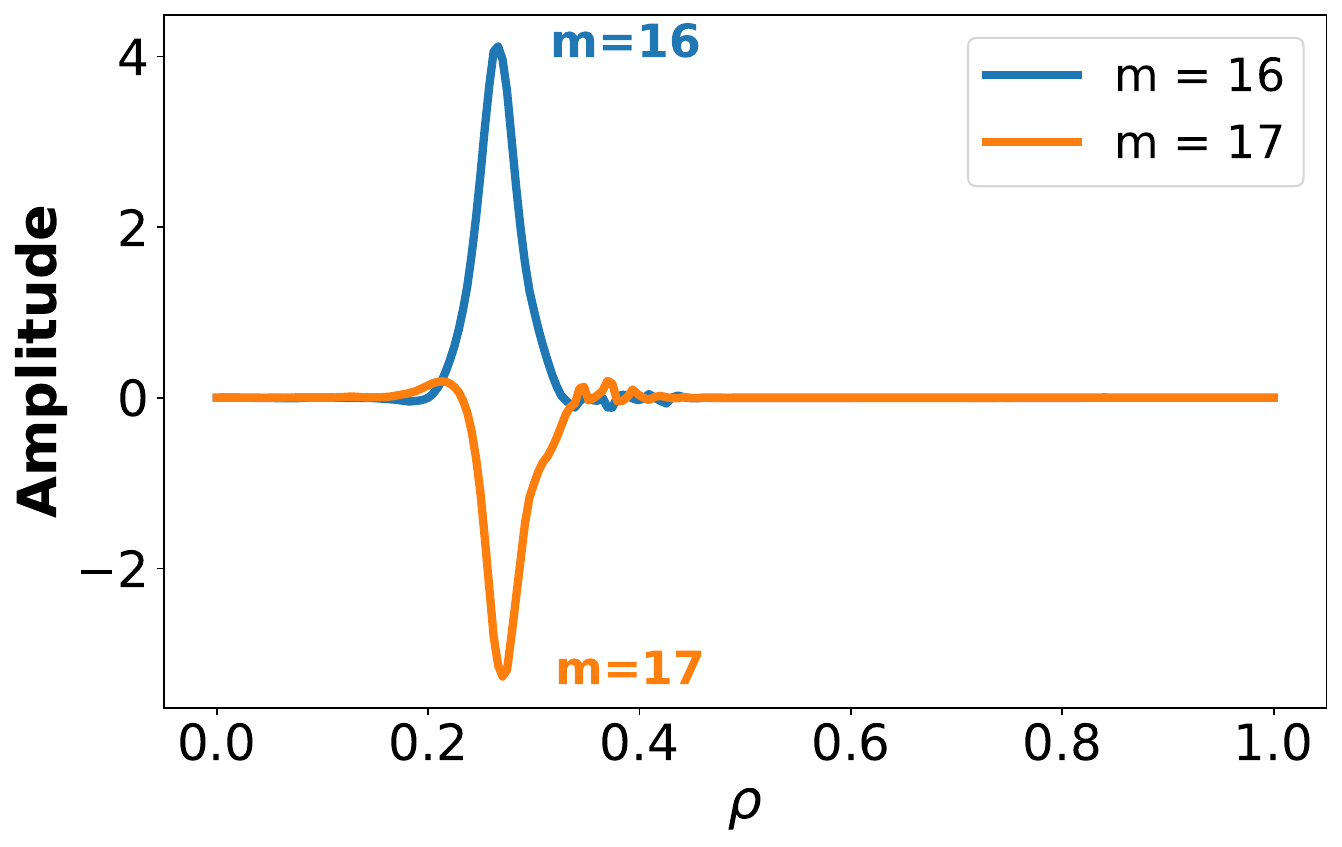}
	}\\[0.6em]

	\subfloat[$\kappa=1.22$ contour \label{con_ba12}]{
		\includegraphics[width=0.42\textwidth,height=5.2cm]{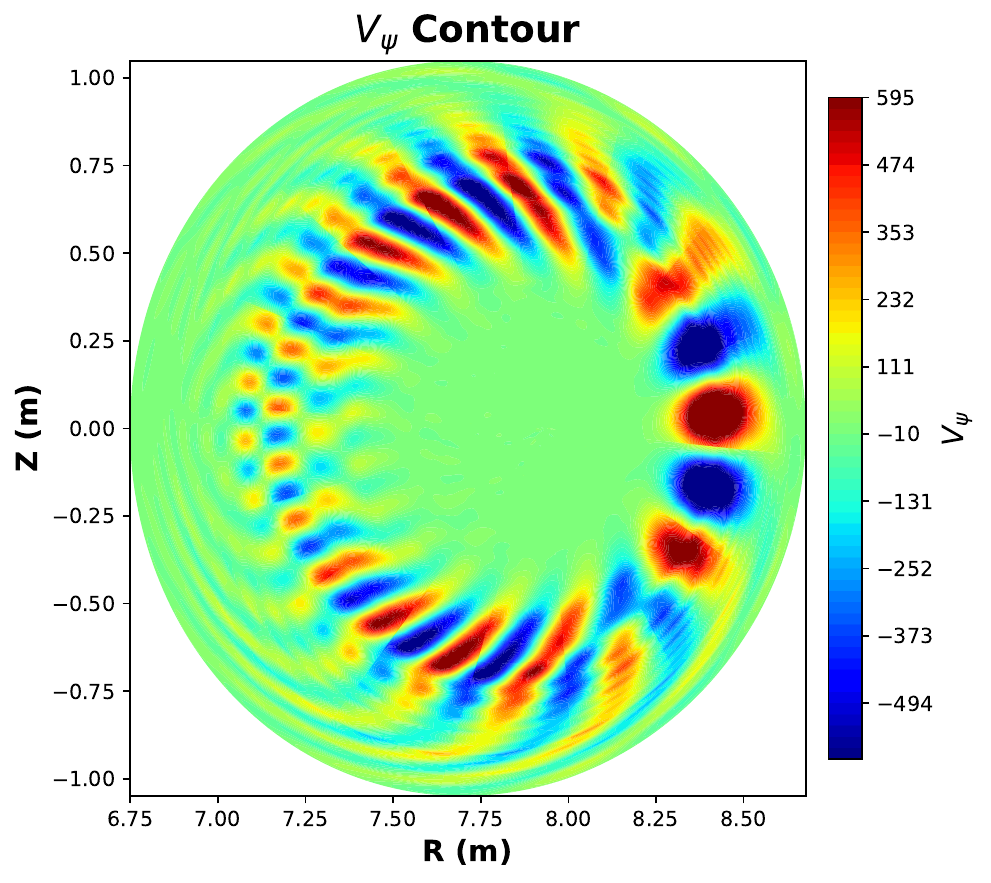}
	}\hfill
	\subfloat[$\kappa=1.22$ PFS \label{pfsba12}]{
		\includegraphics[width=0.42\textwidth,height=5.2cm]{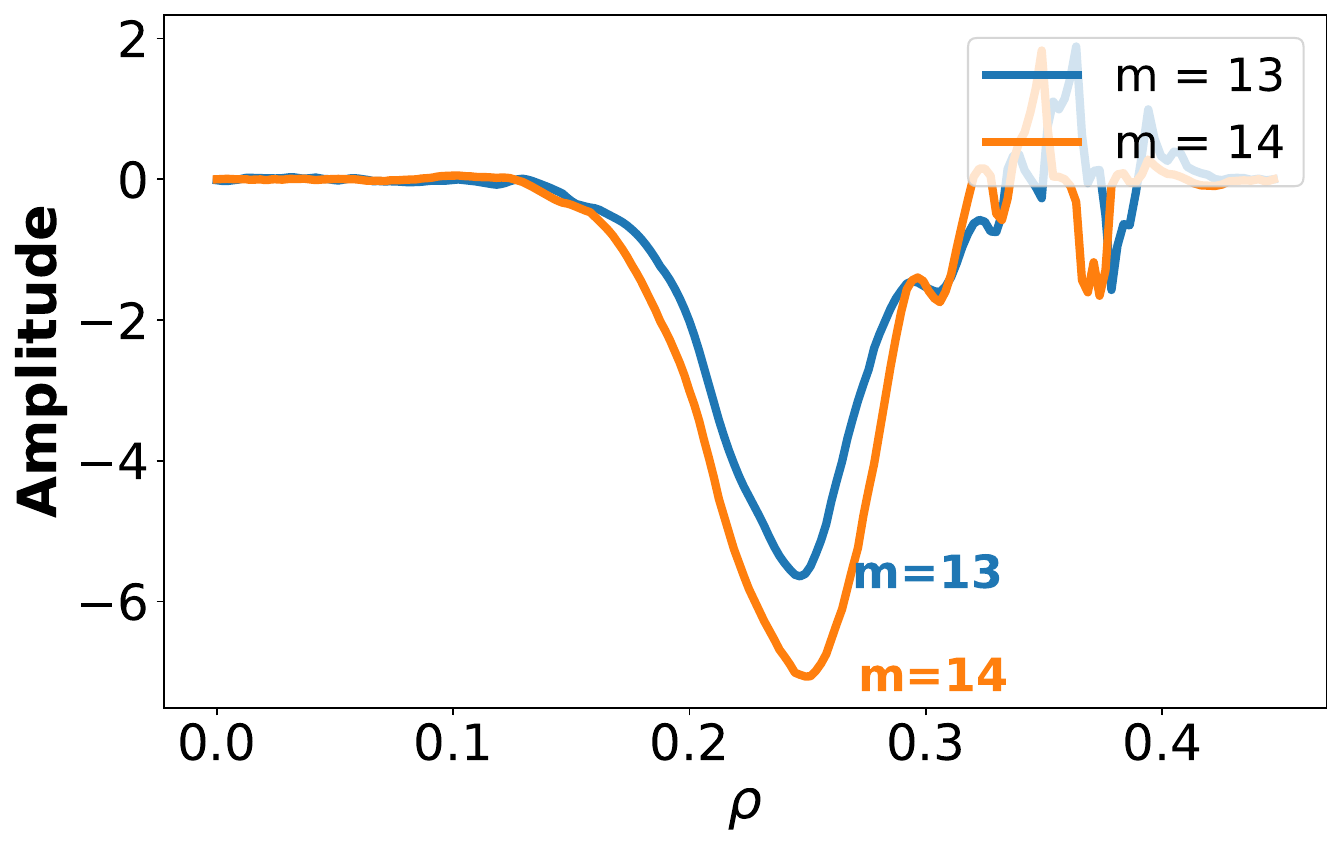}
	}
	
	\caption{
		{Contours of the perturbed normal velocity component and corresponding radial profiles of its {two most dominant} poloidal Fourier components from NIMROD simulations for equilibria with elongations:}
		(a,b) $\kappa=2.0$ (close to the CFETR equilibrium),
		(c,d) $\kappa=1.6$, and
		(e,f) $\kappa=1.22$ (nearly circular).
		{The results show a transition from an odd-parity to an even-parity TAE as the elongation decreases.}
	}
	\label{fig:combined_ba}
\end{figure}
\clearpage

\begin{figure}[H]
	\centering

	\subfloat[$n=1$ PFS ($m=1$--$5$) \label{fig:pfs_n1_m1_5}]{
		\includegraphics[width=0.48\textwidth]{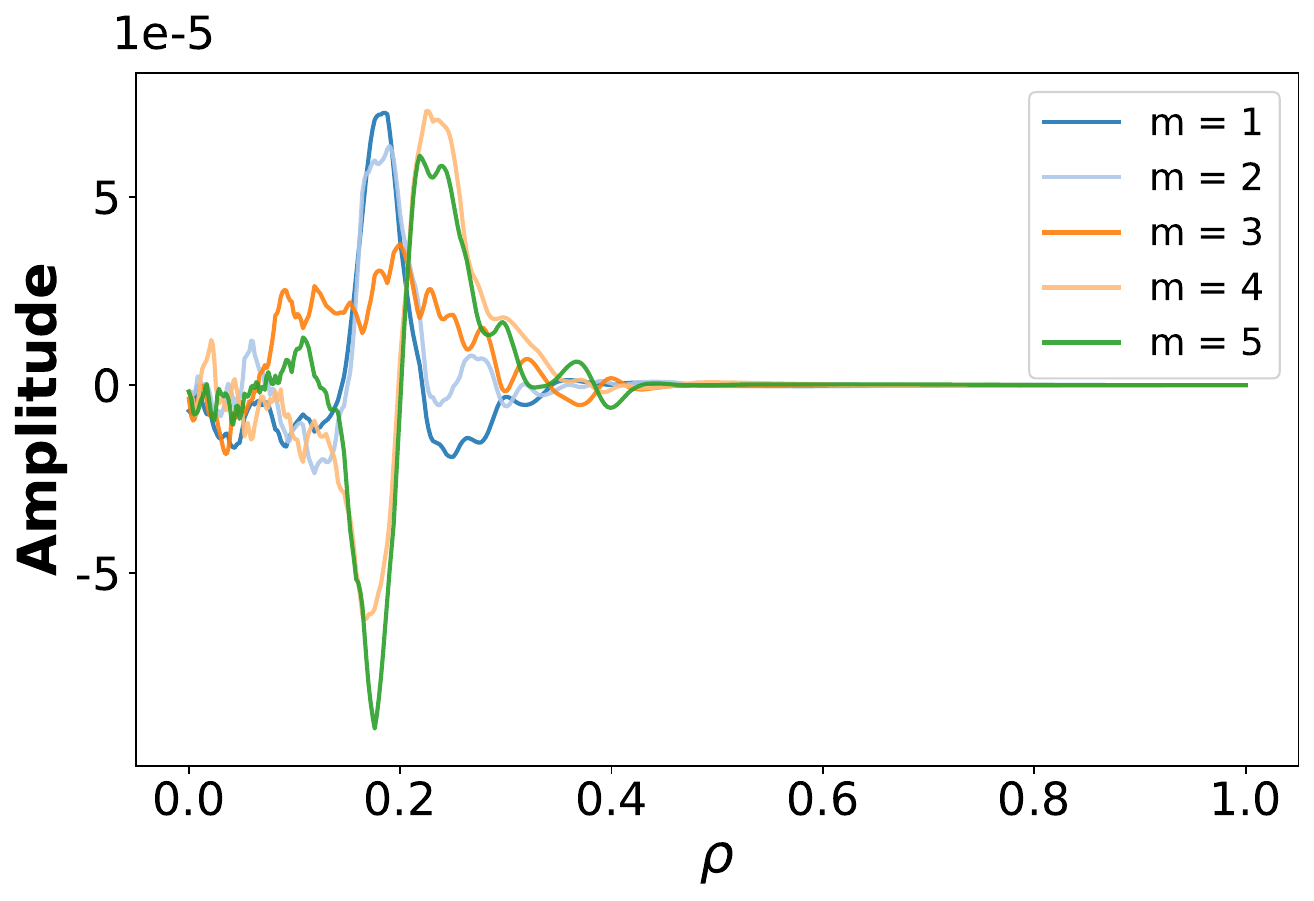}
	}\hfill
	\subfloat[$n=1$ PFS ($m=6$--$9$) \label{fig:pfs_n1_m6_9}]{
		\includegraphics[width=0.48\textwidth]{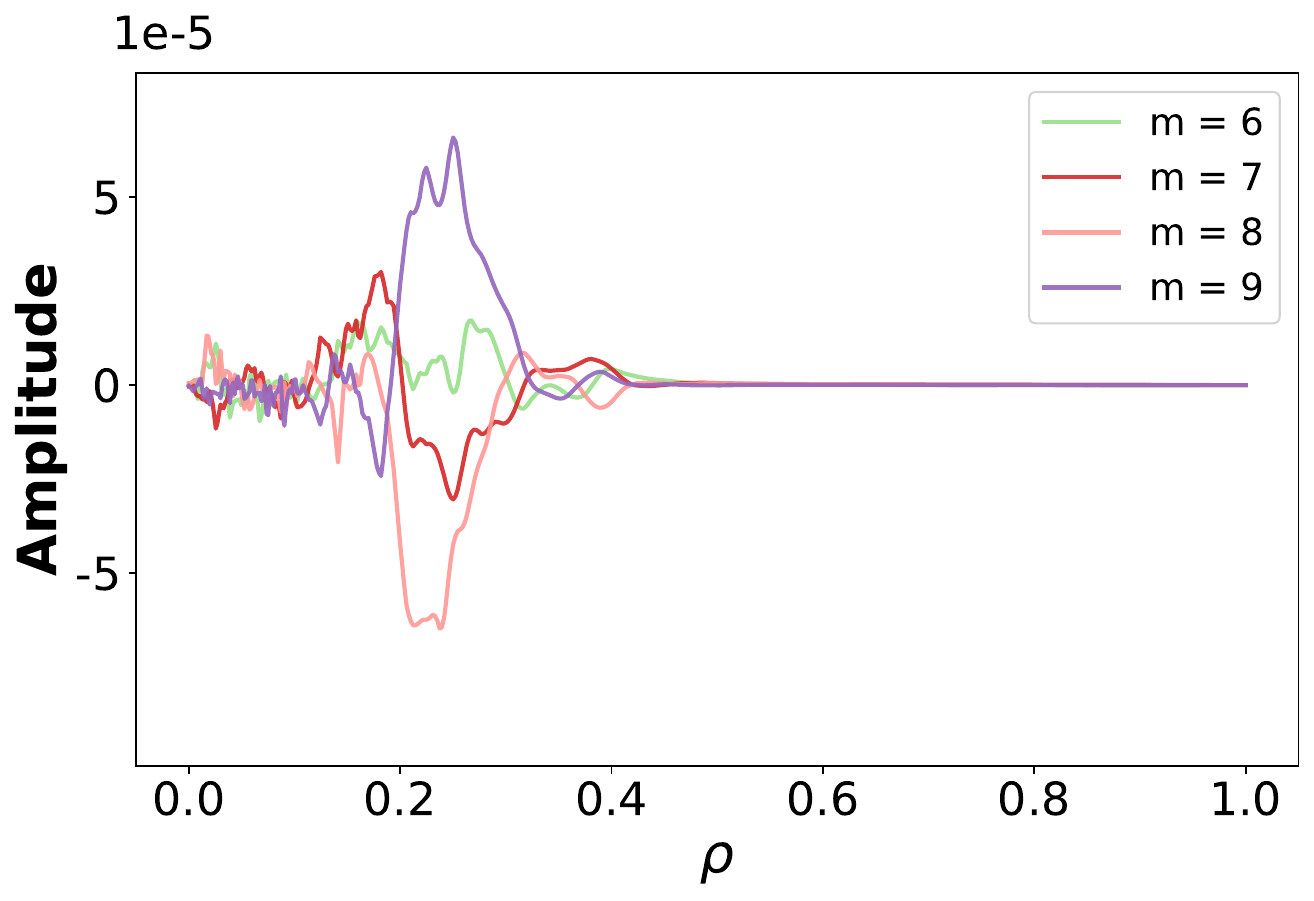}
	}
	
	\vspace{0.6em}

	\subfloat[$n=1$ Alfv\'{e}n continuum \label{fig:continuum_n1}]{
		\includegraphics[width=0.75\textwidth,height=7.0cm]{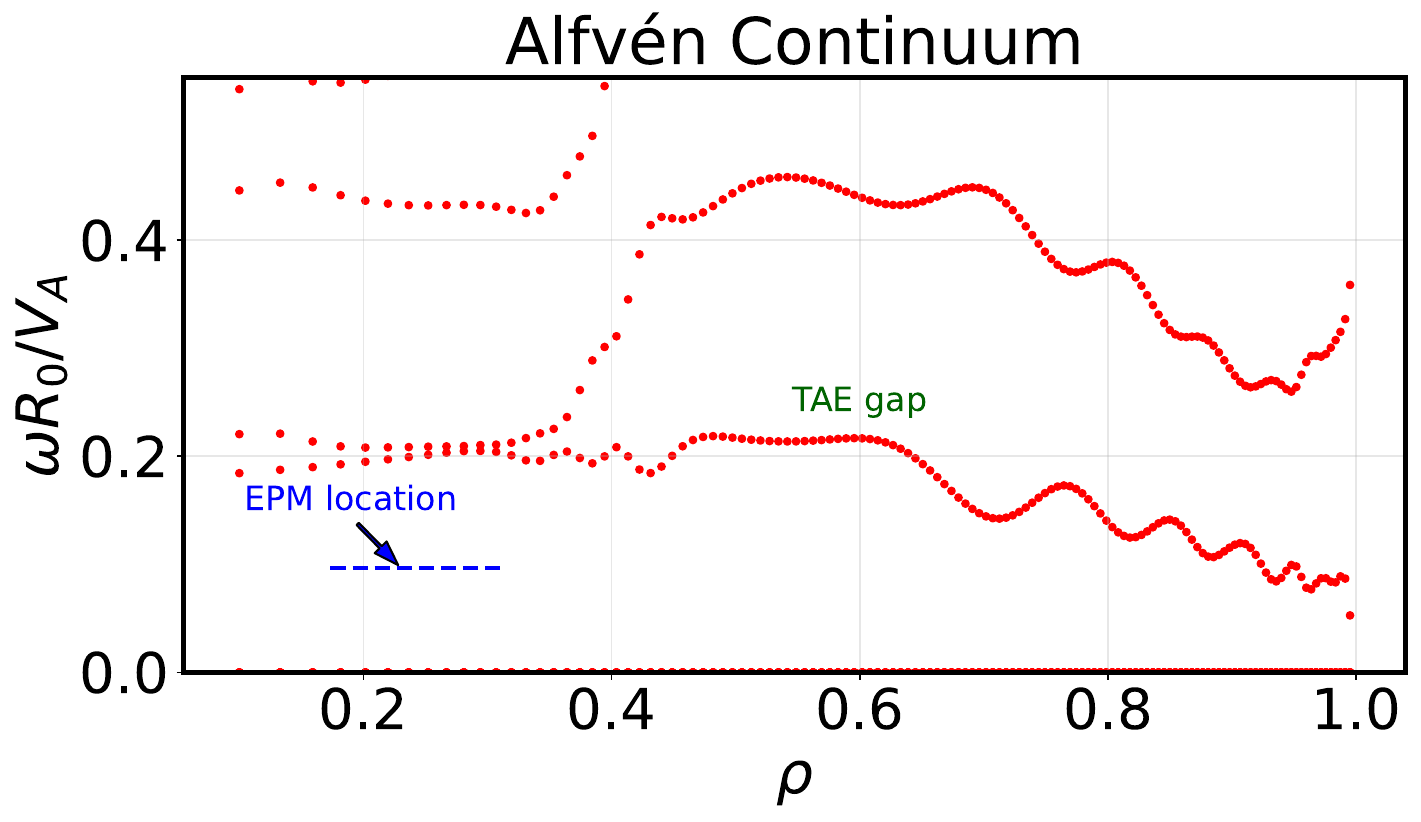}
	}
	
	\caption{
		Radial profiles of the dominant poloidal Fourier components of the $n=1$ mode from the NIMROD simulation,
		shown separately for (a) $m=1$--$5$ and (b) $m=6$--$9$ {components}, and 
		(c) the corresponding Alfv\'{e}n continuum calculated using GTAW.
	}
	\label{fig:combined_pfs_continuum_n1}
\end{figure}
\clearpage

\begin{figure}[H]
	\centering
	
	\subfloat[$\kappa=2.0$ contour \label{fig:pos_k200}]{
		\includegraphics[width=0.435\textwidth,height=5.2cm]{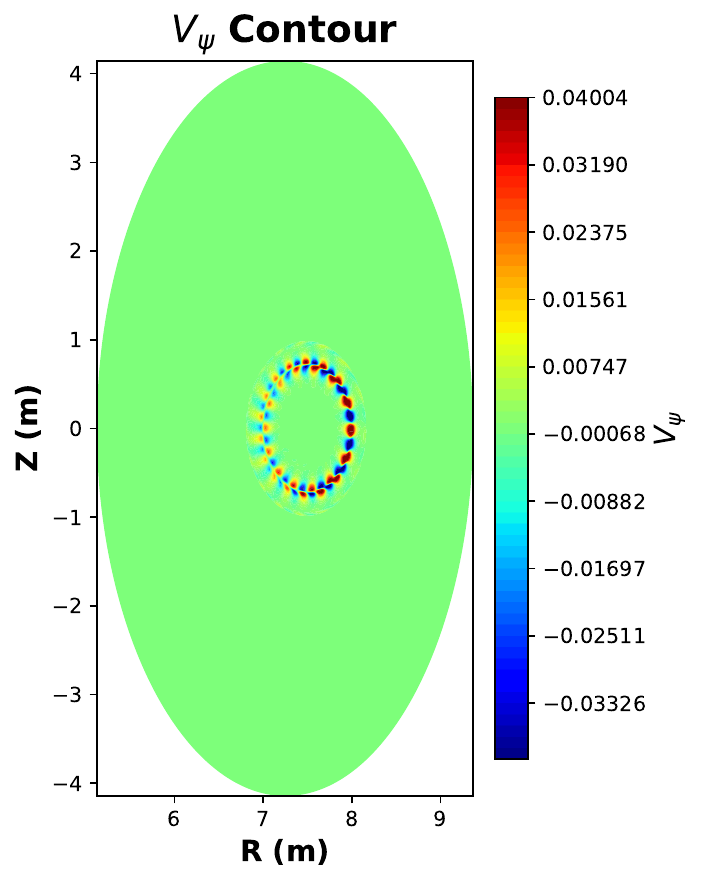}
	}\hfill
	\subfloat[$\kappa=2.0$ PFS \label{pfs_pos_ba20}]{
		\includegraphics[width=0.435\textwidth,height=5.2cm]{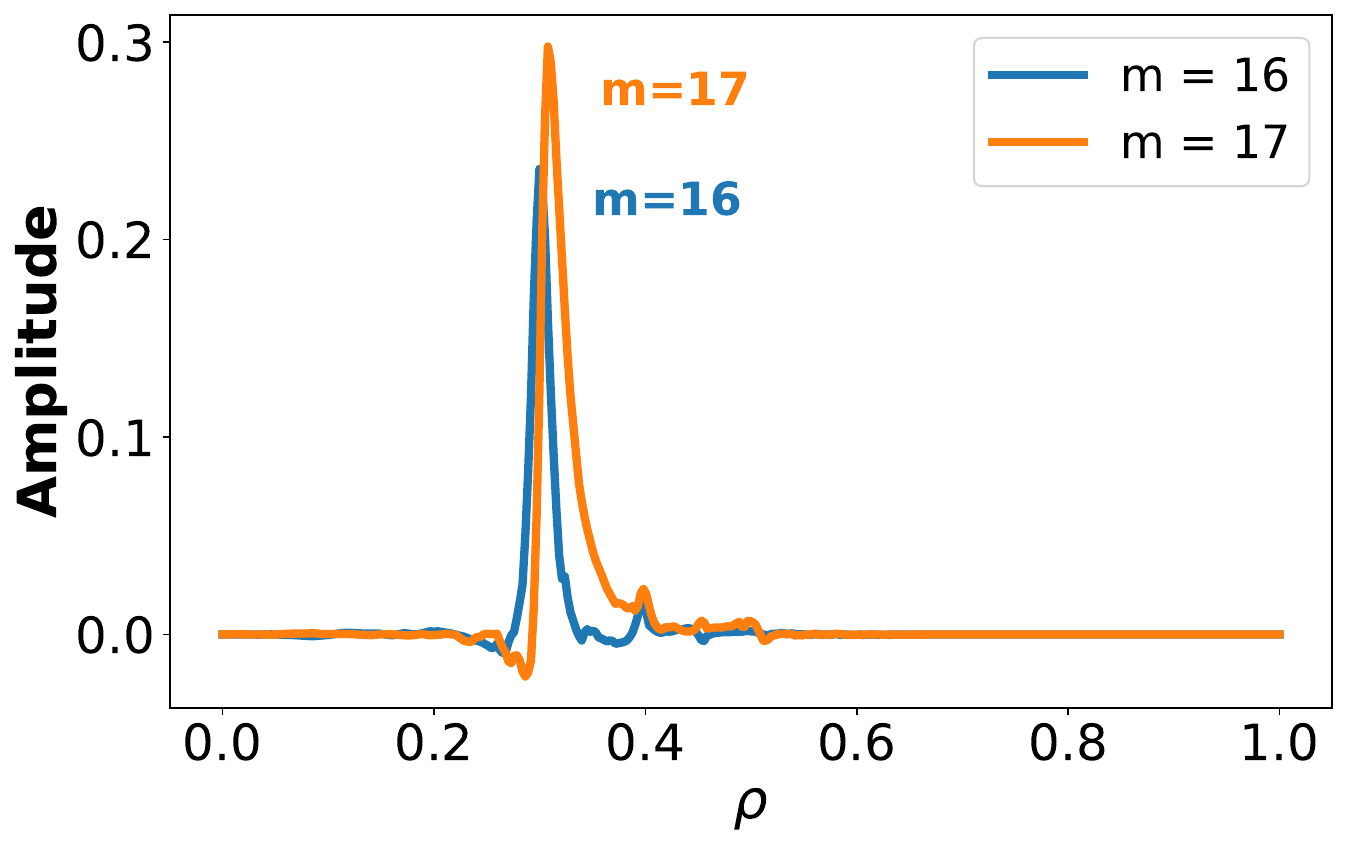}
	}\\[0.6em]

	\subfloat[$\kappa=1.5$ contour \label{fig:pos_k150}]{
		\includegraphics[width=0.42\textwidth,height=5.2cm]{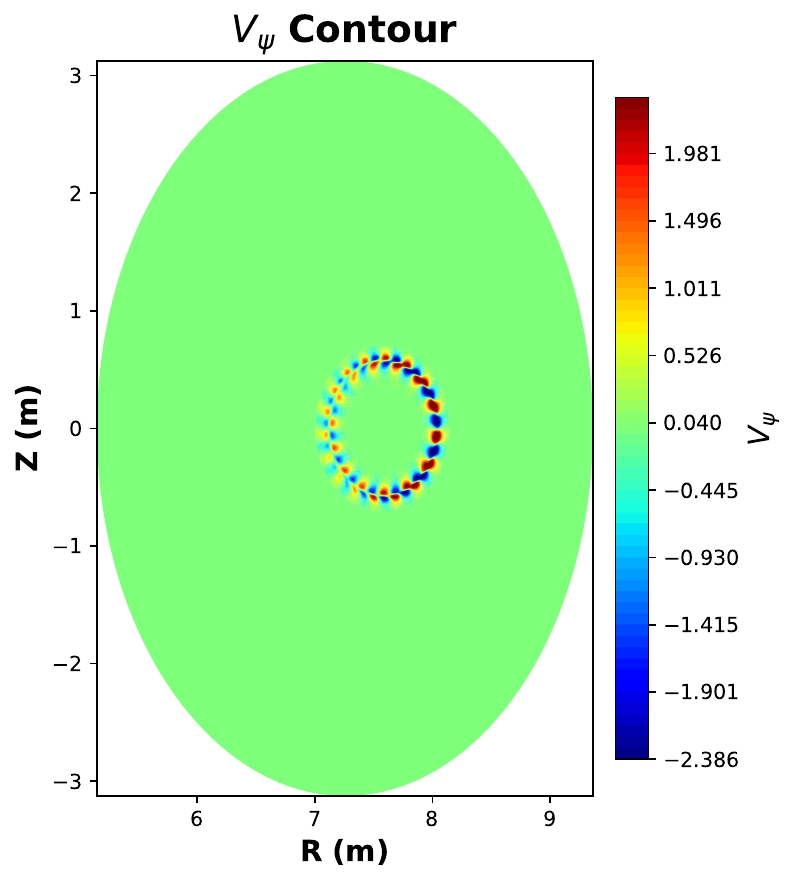}
	}\hfill
	\subfloat[$\kappa=1.5$ PFS \label{pfs_pos_ba12}]{
		\includegraphics[width=0.45\textwidth,height=5.2cm]{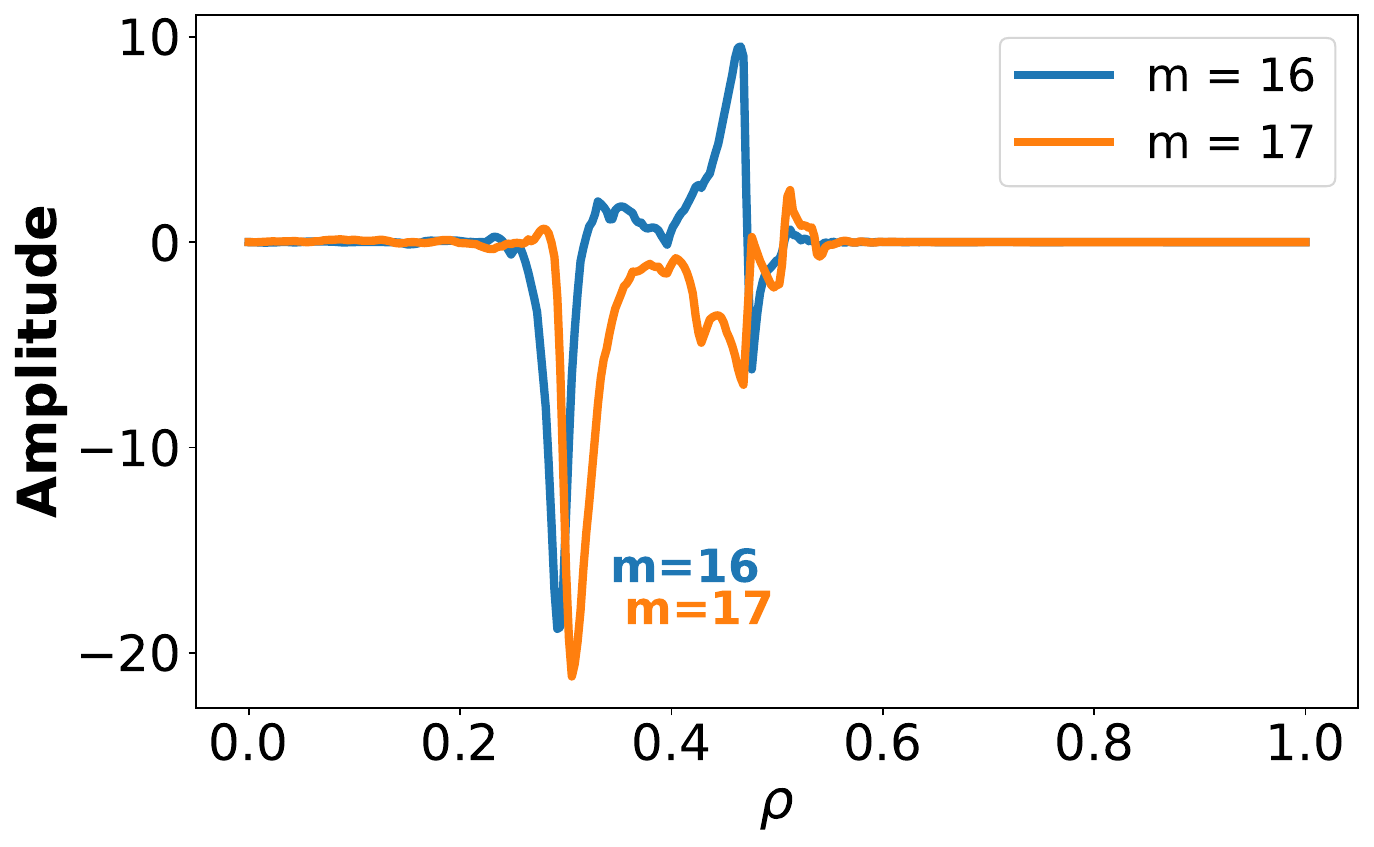}
	}\\[0.6em]

	\subfloat[$\kappa=1.0$ contour \label{fig:pos_k100}]{
		\includegraphics[width=0.42\textwidth,height=5.2cm]{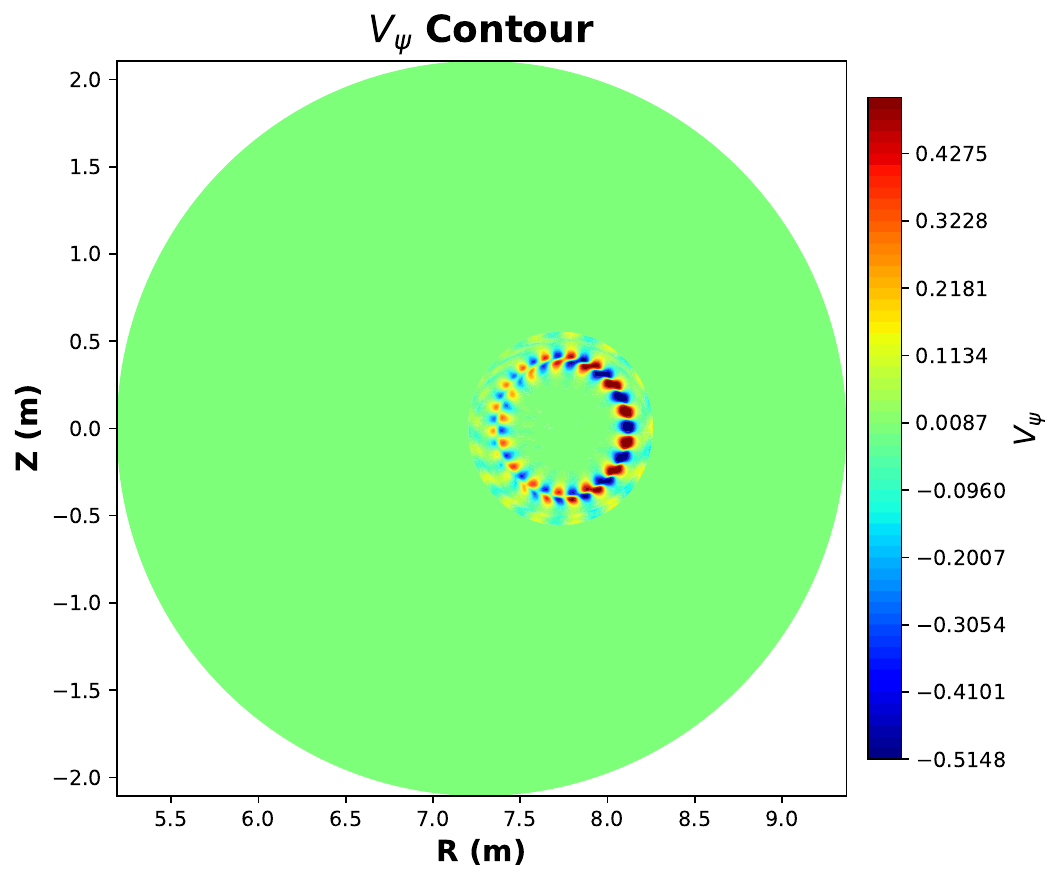}
	}\hfill
	\subfloat[$\kappa=1.0$ PFS \label{pfs_pos_ba11}]{
		\includegraphics[width=0.42\textwidth,height=5.2cm]{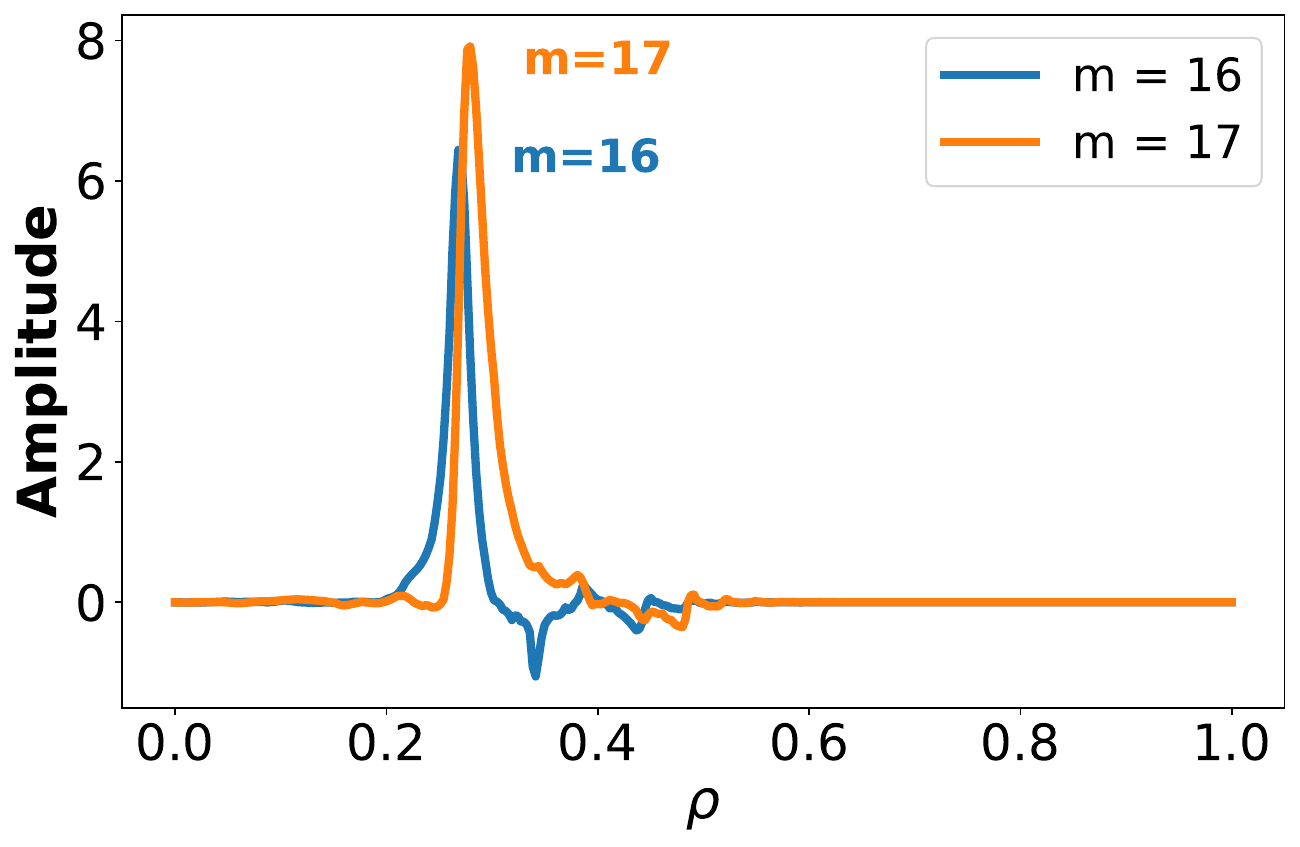}
	}
	
	\caption{
		{{Contours of the perturbed normal velocity component and corresponding radial profiles of its two most dominant poloidal Fourier components from NIMROD simulations for equilibria with elongations:}
			(a,b) $\kappa=2.0$ (close to the CFETR equilibrium),
			(c,d) $\kappa=1.5$, and
			(e,f) $\kappa=1.0$ (circular).
			{These results correspond to equilibria with strong positive magnetic shear.} All modes exhibit ballooning structure with even parity.
	}}
	\label{fig:combined_pshear}
\end{figure}
\clearpage

\section*{References}

\bibliography{sample-1}

\end{document}